\documentclass[twoside]{article}
\usepackage{qic}

\usepackage{amsmath}
\usepackage{amssymb}
\usepackage{braket}
\usepackage{color}

\newtheorem{thm}{Theorem}

\begin{document}
\setlength{\textheight}{8.0truein}    

\runninghead{A limit distribution for a quantum walk driven by a five-diagonal unitary matrix}
            {T. Machida}

\normalsize\textlineskip
\thispagestyle{empty}
\setcounter{page}{1}


\vspace*{0.88truein}

\alphfootnote

\fpage{1}

\centerline{\bf
A limit distribution for a quantum walk driven by a five-diagonal unitary matrix}
\vspace*{0.37truein}
\centerline{\footnotesize
Takuya Machida}
\vspace*{0.015truein}
\centerline{\footnotesize\it College of Industrial Technology, Nihon University, Narashino, Chiba 275-8576, Japan}
\vspace*{0.225truein}

\vspace*{0.21truein}

\abstracts{
In this paper, we work on a quantum walk whose system is manipulated by a five-diagonal unitary matrix, and present long-time limit distributions.
The quantum walk launches off a location and delocalizes in distribution as its system is getting updated.
The five-diagonal matrix contains a phase term and the quantum walk becomes a standard coined walk when the phase term is fixed at special values.
Or, the phase term gives an effect on the quantum walk.
As a result, we will see an explicit form of a long-time limit distribution for a quantum walk driven by the matrix, and thanks to the exact form, we understand how the quantum walker approximately distributes in space after the long-time evolution has been executed on the walk.
}{}{}

\vspace*{10pt}

\keywords{Quantum walk, Limit distribution}
\vspace*{3pt}

\vspace*{1pt}\textlineskip    

\bibliographystyle{qic}

\section{Introduction}

Coined quantum walks, specifically the Hadamard walk, were introduced in~\cite{AharonovDavidovichZagury1993} and their limit law in the case of translationally invariant walks with an arbitrary unitary coin in $U(2)$ are well known, see for instance~\cite{Konno2008}.
There are different methods to derive these results, and a natural one, in view of the assumed translational invariance is the Fourier method, see~\cite{GrimmettJansonScudo2004}.

%
Motivated on limit distributions, we are challenging to discover them analytically one by one.
The limit distribution plays an important roll to describe the behavior of the quantum walk after many steps, and it exactly tells us how the quantum walk behaves depending on its coin operation and the initial state.
To get the concrete representation of limit distributions, we focus on a spatially homogeneous matrix and try to find the limit distribution in this paper so that we precisely understand how the quantum walk is controlled by the coin operation and the initial state.  
%
Quantum walks have been numerically studied and some probability distributions for the walk were visualized.
We consider two cases: in the first case the unitary evolution is given by a matrix and in the second case it is given by the product of a matrix and a simple permutation matrix.
In the first case the limit law is already given in~\cite{Konno2008} and the result is given here in an appendix.
In the second case we obtain a new form of a limit law.
In all cases this is studied by means of Fourier analysis and some numerical examples are included to show the agreement between our theoretical result, namely Theorem~\ref{th:limit_CMV} is section~\ref{sec:limit_CMV} and numerical simulations.

\section{Definition of a quantum walk}
\label{sec:introduction}
A quantum walker with two coin states $\ket{0}$ and $\ket{1}$ is located at points in $\mathbb{Z}=\left\{0,\pm 1,\pm 2,\ldots\right\}$.
The state  of the system is described by a normalized vector on the tensor Hilbert space $\mathcal{H}_p\otimes\mathcal{H}_s$.
The Hilbert space $\mathcal{H}_p$ encodes the integer points and it is spanned by the orthogonal normalized basis $\left\{\ket{x} : x\in\mathbb{Z}\right\}$.
The Hilbert space $\mathcal{H}_s$ represents the coin states and it is spanned by the orthogonal normalized basis $\left\{\ket{0},\ket{1}\right\}$.
The state of the quantum walk at time $t\,(=0,1,2,\ldots)$, represented by $\ket{\Psi_t}\in\mathcal{H}_p\otimes\mathcal{H}_s$, updates with a unitary operation,  
\begin{equation}
 \ket{\Psi_{t+1}}=U\ket{\Psi_t},\label{eq:time_evolution}
\end{equation}
where, given parameters $\rho\in (0,1)$ and $\nu\in\mathbb{R}$,
\begin{align}
 U=&\sum_{x\in\mathbb{Z}}\ket{x-1}\bra{x}\otimes\Bigl(\rho_0^2\ket{1}\bra{0}+\rho_0\overline{\alpha_0}\ket{1}\bra{1}\Bigr)\nonumber\\
 &+\ket{x}\bra{x}\otimes\Bigl\{-\alpha_0\rho_0\ket{0}\bra{0}-(1-\rho_0^2)\ket{0}\bra{1}-(1-\rho_0^2)\ket{1}\bra{0}+\rho_0\overline{\alpha_0}\ket{1}\bra{1}\Bigr\}\nonumber\\
 &+\ket{x+1}\bra{x}\otimes\Bigl(-\alpha_0\rho_0\ket{0}\bra{0}+\rho_0^2\ket{0}\bra{1}\Bigr),\nonumber\\
 =&\left\{\sum_{x\in\mathbb{Z}}\Bigl(\ket{x}\bra{x}\otimes\ket{0}\bra{0}+\ket{x-1}\bra{x}\otimes\ket{1}\bra{1}\Bigr)\right\}\nonumber\\
 &\times\left\{\sum_{x\in\mathbb{Z}}\ket{x}\bra{x}\otimes\Bigl\{\rho_0 e^{i\nu/2}\ket{0}\bra{0}+\rho e^{i\nu/2}\ket{0}\bra{1}+\rho e^{-i\nu/2}\ket{1}\bra{0}-\rho_0 e^{-i\nu/2}\ket{1}\bra{1}\Bigr\}\right\}\nonumber\\
 &\times\left\{\sum_{x\in\mathbb{Z}}\Bigl(\ket{x+1}\bra{x}\otimes\ket{0}\bra{0}+\ket{x}\bra{x}\otimes\ket{1}\bra{1}\Bigr)\right\}\nonumber\\
 &\times\left\{\sum_{x\in\mathbb{Z}}\ket{x}\bra{x}\otimes\Bigl\{-\rho e^{i\nu/2}\ket{0}\bra{0}+\rho_0 e^{-i\nu/2}\ket{0}\bra{1}-\rho_0 e^{i\nu/2}\ket{1}\bra{0}-\rho e^{-i\nu/2}\ket{1}\bra{1}\Bigr\}\right\},
\end{align}
with $\alpha_0=\rho\, e^{i\nu}$ and $\rho_0=\sqrt{1-\rho^2}$.
The notation $i$ denotes the imaginary unit in complex numbers.
The operation $U$ contains a phase parameter $\nu$, differently from the operations in~\cite{MachidaKonno2010}.
We assume $\ket{\Psi_0}=\ket{0}\otimes\left(\alpha\ket{0}+\beta\ket{1}\right)\,(=\ket{0}\otimes\ket{\phi})$ with $|\alpha|^2+|\beta|^2=1$.

The unitary operation $U$ can be decomposed to the product of two unitary operations, $U=VU_{f}$ with
\begin{align}
 V=&\sum_{x\in\mathbb{Z}}\ket{x-1}\bra{x}\otimes\Bigl(\rho_0\overline{\alpha_0}\ket{1}\bra{0}+\rho_0^2\ket{1}\bra{1}\Bigr)\nonumber\\
 &+\ket{x}\bra{x}\otimes\Bigl\{-(1-\rho_0^2)\ket{0}\bra{0}-\alpha_0\rho_0\ket{0}\bra{1}+\rho_0\overline{\alpha_0}\ket{1}\bra{0}-(1-\rho_0^2)\ket{1}\bra{1}\Bigr\}\nonumber\\
 &+\ket{x+1}\bra{x}\otimes\Bigl(\rho_0^2\ket{0}\bra{0}-\alpha_0\rho_0\ket{0}\bra{1}\Bigr),\\
 U_{f}=&\sum_{x\in\mathbb{Z}}\ket{x}\bra{x}\otimes\Bigl(\ket{0}\bra{1}+\ket{1}\bra{0}\Bigr).
\end{align}
Once standard bases are given to the Hilbert spaces $\mathcal{H}_p$ and $\mathcal{H}_c$, we get a matrix representation of $V$,
\begin{equation}
 V=\begin{bmatrix}
    & \vdots & \vdots & \vdots & \vdots & \vdots & \vdots & \\
    \cdots & 0& 0& 0 & 0 & 0 & 0 & \cdots\\
    \cdots & \rho_0\overline{\alpha_0} & \rho_0^2 & 0 & 0 & 0 & 0 & \cdots\\
    \cdots & -(1-\rho_0^2) & -\alpha_0\rho_0 & 0 & 0 & 0 & 0 & \cdots\\
    \cdots & \rho_0\overline{\alpha_0} & -(1-\rho_0^2) & \rho_0\overline{\alpha_0} & \rho_0^2 & 0 & 0 & \cdots\\
    \cdots & \rho_0^2 & -\alpha_0\rho_0 & -(1-\rho_0^2) & -\alpha_0\rho_0 & 0 & 0 & \cdots\\
    \cdots & 0 & 0 & \rho_0\overline{\alpha_0} & -(1-\rho_0^2) & \rho_0\overline{\alpha_0} & \rho_0^2 & \cdots\\
    \cdots & 0 & 0 & \rho_0^2 & -\alpha_0\rho_0 & -(1-\rho_0^2) & -\alpha_0\rho_0 & \cdots\\
     \cdots & 0 & 0 & 0 & 0 & \rho_0\overline{\alpha_0} & -(1-\rho_0^2) & \cdots\\
     \cdots & 0 & 0 & 0 & 0 & \rho_0^2 & -\alpha_0\rho_0 & \cdots\\
    \cdots & 0 & 0 & 0 & 0 & 0 & 0 & \cdots\\
    & \vdots & \vdots & \vdots & \vdots & \vdots & \vdots & 
   \end{bmatrix},
\end{equation}
which belongs to a class of CMV matrices~\cite{CanteroMoralGrunbaumVelazquez2010}.
The operation $U_f$ is expressed as
\begin{equation}
 U_f=\begin{bmatrix}
    & \vdots & \vdots & \vdots & \vdots & \vdots & \vdots & \\
    \cdots & 0 & 1 & 0 & 0 & 0 & 0 & \cdots\\
    \cdots & 1 & 0 & 0 & 0 & 0 & 0 & \cdots\\
    \cdots & 0 & 0 & 0 & 1 & 0 & 0 & \cdots\\
    \cdots & 0 & 0 & 1 & 0 & 0 & 0 & \cdots\\
    \cdots & 0 & 0 & 0 & 0 & 0 & 1 & \cdots\\
     \cdots & 0 & 0 & 0 & 0 & 1 & 0 & \cdots\\
    & \vdots & \vdots & \vdots & \vdots & \vdots & \vdots & 
   \end{bmatrix},
\end{equation}
and the unitary operation $U$ becomes
\begin{equation}
 U=\begin{bmatrix}
    & \vdots & \vdots & \vdots & \vdots & \vdots & \vdots & \\
    \cdots & 0& 0& 0 & 0 & 0 & 0 & \cdots\\
    \cdots & \rho_0^2 & \rho_0\overline{\alpha_0} & 0 & 0 & 0 & 0 & \cdots\\
    \cdots & -\alpha_0\rho_0 & -(1-\rho_0^2) & 0 & 0 & 0 & 0 & \cdots\\
    \cdots & -(1-\rho_0^2) & \rho_0\overline{\alpha_0} & \rho_0^2 & \rho_0\overline{\alpha_0} & 0 & 0 & \cdots\\
    \cdots & -\alpha_0\rho_0 & \rho_0^2 & -\alpha_0\rho_0 & -(1-\rho_0^2) & 0 & 0 & \cdots\\
    \cdots & 0 & 0 & -(1-\rho_0^2) & \rho_0\overline{\alpha_0} & \rho_0^2 & \rho_0\overline{\alpha_0} & \cdots\\
    \cdots & 0 & 0 & -\alpha_0\rho_0 & \rho_0^2 & -\alpha_0\rho_0 & -(1-\rho_0^2) & \cdots\\
     \cdots & 0 & 0 & 0 & 0 & -(1-\rho_0^2) & \rho_0\overline{\alpha_0} & \cdots\\
     \cdots & 0 & 0 & 0 & 0 & -\alpha_0\rho_0 & \rho_0^2 & \cdots\\
    \cdots & 0 & 0 & 0 & 0 & 0 & 0 & \cdots\\
    & \vdots & \vdots & \vdots & \vdots & \vdots & \vdots & 
   \end{bmatrix}.
\end{equation}

With $\bigl|\bigl|\,\ket{\psi}\,\bigr|\bigr|=\sqrt{\braket{\psi|\psi}}\quad(\,\ket{\psi}\in\mathcal{H}_s\,)$, the quantum walker is observed at position $x\in\mathbb{Z}$ at time $t\in\left\{0,1,2,\ldots\right\}$ with probability
\begin{align}
 \mathbb{P}(X_t=x)
 =&\bra{\Psi_t}\Bigl\{\ket{x}\bra{x}\otimes\bigl(\,\ket{0}\bra{0}+\ket{1}\bra{1}\,\bigr)\Bigr\}\ket{\Psi_t}\nonumber\\
 =&\biggl|\biggl|\,\Bigl\{\bra{x}\otimes\bigl(\,\ket{0}\bra{0}+\ket{1}\bra{1}\,\bigr)\Bigr\}\ket{\Psi_t}\,\biggr|\biggr|^2.
\end{align}

\section{Fourier transform}
The Fourier analysis has been used for discovering the interesting behavior of quantum walks~\cite{Machida2016a}.
Let $\ket{\hat\psi_t(k)}\,\in\mathbb{C}^2\,(k\in [-\pi,\pi))$ be the Fourier transform of the quantum walk at time $t$,
\begin{equation}
 \ket{\hat\psi_t(k)}=\sum_{x\in\mathbb{Z}}e^{-ikx}\Bigl\{\bra{x}\otimes\bigl(\,\ket{0}\bra{0}+\ket{1}\bra{1}\,\bigr)\Bigr\}\ket{\Psi_t},
\end{equation}
from which the Fourier inverse transform reproduces the system of quantum walk,
\begin{equation}
 \ket{\Psi_t}=\sum_{x\in\mathbb{Z}}\ket{x}\otimes\int_{-\pi}^\pi e^{ikx}\ket{\hat\psi_t(k)}\,\frac{dk}{2\pi}.
\end{equation}
Once assigning a standard basis to the Hilbert space $\mathcal{H}_s$,
\begin{equation}
 \ket{0}=\begin{bmatrix}
	  1\\0
	 \end{bmatrix},\quad
	 \ket{1}=\begin{bmatrix}
		  0\\1
		 \end{bmatrix}.
\end{equation}
we can tell from Eq.~\eqref{eq:time_evolution} that the Fourier transform gets updated with a $2\times 2$ unitary matrix $\hat{U}(k)$,    
\begin{equation}
 \ket{\hat\psi_{t+1}(k)}=\hat{U}(k)\ket{\hat\psi_t(k)},\label{eq:time_evolution_FT}
\end{equation}
where
\begin{equation}
 \hat{U}(k)=R\left(-\frac{\nu}{2}\right)H(k) R\left(\frac{\nu}{2}\right),
\end{equation}
with
\begin{align}
 R(\varphi)=&\begin{bmatrix}
	 e^{i\varphi} & 0\\
	 0 & e^{-i\varphi}
	\end{bmatrix}\quad (\varphi\in\mathbb{R}),\\
 H(k)=&\begin{bmatrix}
	-\rho\rho_0(e^{i\nu}+e^{-i(k-\nu)}) & -\rho^2e^{i\nu}+\rho_0^2 e^{-i(k-\nu)}\\
	-\rho^2e^{-i\nu}+\rho_0^2 e^{i(k-\nu)} & \rho\rho_0(e^{-i\nu}+e^{i(k-\nu)})
       \end{bmatrix}.
\end{align}

\begin{align}
 \ket{\hat\psi_t(k)}
 =\hat{U}(k)^t\ket{\hat\psi_0(k)}
 =&R\left(-\frac{\nu}{2}\right)H(k)^t R\left(\frac{\nu}{2}\right)\ket{\hat\psi_0(k)}\nonumber\\
 =&R\left(-\frac{\nu}{2}\right)H(k)^t \ket{\tilde\phi},
\end{align}
where, remembering $\ket{\phi}=\alpha\ket{0}+\beta\ket{1}$, we get
\begin{equation}
 \ket{\tilde\phi}
 =R\left(\frac{\nu}{2}\right)\ket{\hat\psi_0(k)}
 =R\left(\frac{\nu}{2}\right)\ket{\phi}
 =\begin{bmatrix}
    e^{i\nu/2}\alpha\\[1mm]
    e^{-i\nu/2}\beta
 \end{bmatrix}.
\end{equation}
Note that $|e^{i\nu/2}\alpha|^2+|e^{-i\nu/2}\beta|^2=|\alpha|^2+|\beta|^2=1$.

The Fourier inverse transform
\begin{equation}
 \ket{\Psi_t}=\sum_{x\in\mathbb{Z}}\ket{x}\otimes R\left(-\frac{\nu}{2}\right)\int_{-\pi}^\pi e^{ikx} H(k)^t \ket{\tilde\phi}\,\frac{dk}{2\pi},
\end{equation}
gives another representation of the probability distribution on Fourier space,
\begin{align}
 \mathbb{P}(X_t=x)
 =&\left|\left|\, R\left(-\frac{\nu}{2}\right)\int_{-\pi}^\pi e^{ikx} H(k)^t \ket{\tilde\phi}\,\frac{dk}{2\pi}\,\right|\right|^2\nonumber\\
 =&\left|\left|\, \int_{-\pi}^\pi e^{ikx} H(k)^t \ket{\tilde\phi}\,\frac{dk}{2\pi}\,\right|\right|^2\nonumber\\
 =&\left|\left|\, \int_{-\pi-\nu}^{\pi-\nu} e^{i(k+\nu)x} H(k+\nu)^t \ket{\tilde\phi}\,\frac{dk}{2\pi}\,\right|\right|^2\nonumber\\
 =&\left|\left|\,  \int_{-\pi}^{\pi} e^{ikx}  \tilde{H}(k)^t \ket{\tilde\phi}\,\frac{dk}{2\pi}\,\right|\right|^2,
\end{align}
with
\begin{equation}
 \tilde{H}(k)=H(k+\nu)=\begin{bmatrix}
			-\rho\rho_0(e^{i\nu}+e^{-ik}) & -\rho^2e^{i\nu}+\rho_0^2 e^{-ik}\\
			-\rho^2e^{-i\nu}+\rho_0^2 e^{ik} & \rho\rho_0(e^{-i\nu}+e^{ik})
		       \end{bmatrix}.
\end{equation}
We are, hence, allowed to analyze the quantum walk defined by
\begin{equation}
 \ket{\hat\psi_t(k)}=\tilde{H}(k)^t \ket{\tilde\phi},\label{eq:time_evolution_alternateFT}
\end{equation}
as long as we focus on the probability distribution.
After this point, we are going to concentrate on Eq.~\eqref{eq:time_evolution_alternateFT} instead of Eq.~\eqref{eq:time_evolution_FT}.

\section{Limit distribution}
In this section we see a limit distribution which catches the features of quantum walk.
It is  separately introduced for $(\rho, \nu)= (1/\sqrt{2}, \pi/2+n\pi)\,(n\in\mathbb{Z})$ and for $(\rho, \nu)\neq (1/\sqrt{2}, \pi/2+n\pi)\,(n\in\mathbb{Z})$ because the quantum walk for the first case is equivalent to the standard coined quantum walk whose limit distribution was already proved~\cite{Konno2008}.
The statement of the limit distribution for the second case, however, contained the one for the first case.
\subsection{$(\rho, \nu)= (1/\sqrt{2}, \pi/2+n\pi)\quad(n\in\mathbb{Z})$}

The operation $\tilde{H}(k)$ contains a 2-step evolution of a standard coined quantum walk,
\begin{align}
 \tilde{H}(k)
=&-i\,(-1)^n \left\{R\left(-\frac{k}{2}\right)\cdot
 \frac{1}{\sqrt{2}}
 \begin{bmatrix}
  e^{-(-1)^n \,i\,\pi/4} & -e^{-(-1)^n \, i\,\pi/4}\\[2mm]
  -e^{(-1)^n \, i\,\pi/4} & -e^{(-1)^n \, i\,\pi/4}
 \end{bmatrix}\right\}^2\nonumber\\
=&-i\,(-1)^n \left\{R\left(-\frac{k}{2}\right)V\right\}^2,
\end{align}
where
\begin{equation}
 V=\frac{1}{\sqrt{2}}
  \begin{bmatrix}
   e^{-(-1)^n \, i\,\pi/4} & -e^{-(-1)^n \, i\,\pi/4}\\[2mm]
   -e^{(-1)^n \, i\,\pi/4} & -e^{(-1)^n \, i\,\pi/4}
  \end{bmatrix}.
\end{equation}
We should note that
\begin{equation}
 \left\{R\left(-\frac{k}{2}\right)V\right\}^2
  =\frac{1}{2}
  \begin{bmatrix}
    1-(-1)^n \,i\,e^{-ik} & 1+(-1)^n \,i\,e^{-ik}\\[2mm]
    -1+(-1)^n \,i\,e^{ik} & 1+(-1)^n \,i\,e^{ik}
  \end{bmatrix}.
\end{equation}
We, hence, found the walker at position $x$ at time $t$ with probability
\begin{align}
 \mathbb{P}(X_t=x)
 =&\left|\left|\,  \int_{-\pi}^{\pi} e^{ikx}
 \cdot (-i)^t\,(-1)^{nt} \left\{R\left(-\frac{k}{2}\right)V\right\}^{2t}
 \ket{\tilde\phi}\,\frac{dk}{2\pi}\,\right|\right|^2\nonumber\\
 =&\left|\left|\,  \int_{-\pi}^{\pi} e^{ikx}
 \left\{R\left(-\frac{k}{2}\right)V\right\}^{2t}
 \ket{\tilde\phi}\,\frac{dk}{2\pi}\,\right|\right|^2.
\end{align}
The quantum walk can be analyzed in a similar way to the standard coined walk, and we get a limit distribution, whose proof is omitted here because the readers refer to the method to compute it by Fourier analysis in~\cite{Machida2016a}. 

\bigskip

Assumed $(\rho, \nu)=(1/\sqrt{2}, \pi/2+n\pi)\,(n\in\mathbb{Z})$ and given the localized initial state of the form $\ket{\Psi_0}=\ket{0}\otimes\left(\alpha\ket{0}+\beta\ket{1}\right)$\,($\alpha,\beta\in\mathbb{C}$ such that $|\alpha|^2+|\beta|^2=1$), the quantum walker converges in distribution.
For any real number $x$, we have
\begin{equation}
 \lim_{t\to\infty}\mathbb{P}\left(\frac{X_t}{t}\leq x\right)
  =\int_{-\infty}^x \frac{1}{\pi(1-y^2)\sqrt{1-2y^2}}\,\Theta_n(y)\,I_{(-1/\sqrt{2},\, 1/\sqrt{2}\,)}(y)\,dy,\label{eq:limit_standard_QW}
\end{equation}
where
\begin{align}
 \Theta_n(x)=& 1+\Bigl\{|\alpha|^2+|\beta|^2+(-1)^n \cdot 2\Im(\alpha\overline{\beta})\Bigr\}x,\\
 I_{(-1/\sqrt{2},\, 1/\sqrt{2}\,)}(x)=&\left\{\begin{array}{cl}
					1&(x\in (-1/\sqrt{2},\, 1/\sqrt{2}\,))\\
					       0&(x\notin (-1/\sqrt{2},\, 1/\sqrt{2}\,))
					      \end{array}\right..
\end{align}
The notation $\Im(z)$ denotes the imaginary part of a complex number $z$.

\bigskip

\begin{figure}[h]
\begin{center}
\begin{minipage}{50mm}
  \begin{center}
  (a) $(\alpha, \beta) = (1/\sqrt{2}\,,i/\sqrt{2}\,)$
  \end{center}
 \end{minipage}
 \begin{minipage}{50mm}
  \begin{center}
  (b) $(\alpha, \beta) = (1,0)$
  \end{center}
 \end{minipage}\\
 \bigskip
 \begin{minipage}{50mm}
  \begin{center}
   \includegraphics[scale=0.4]{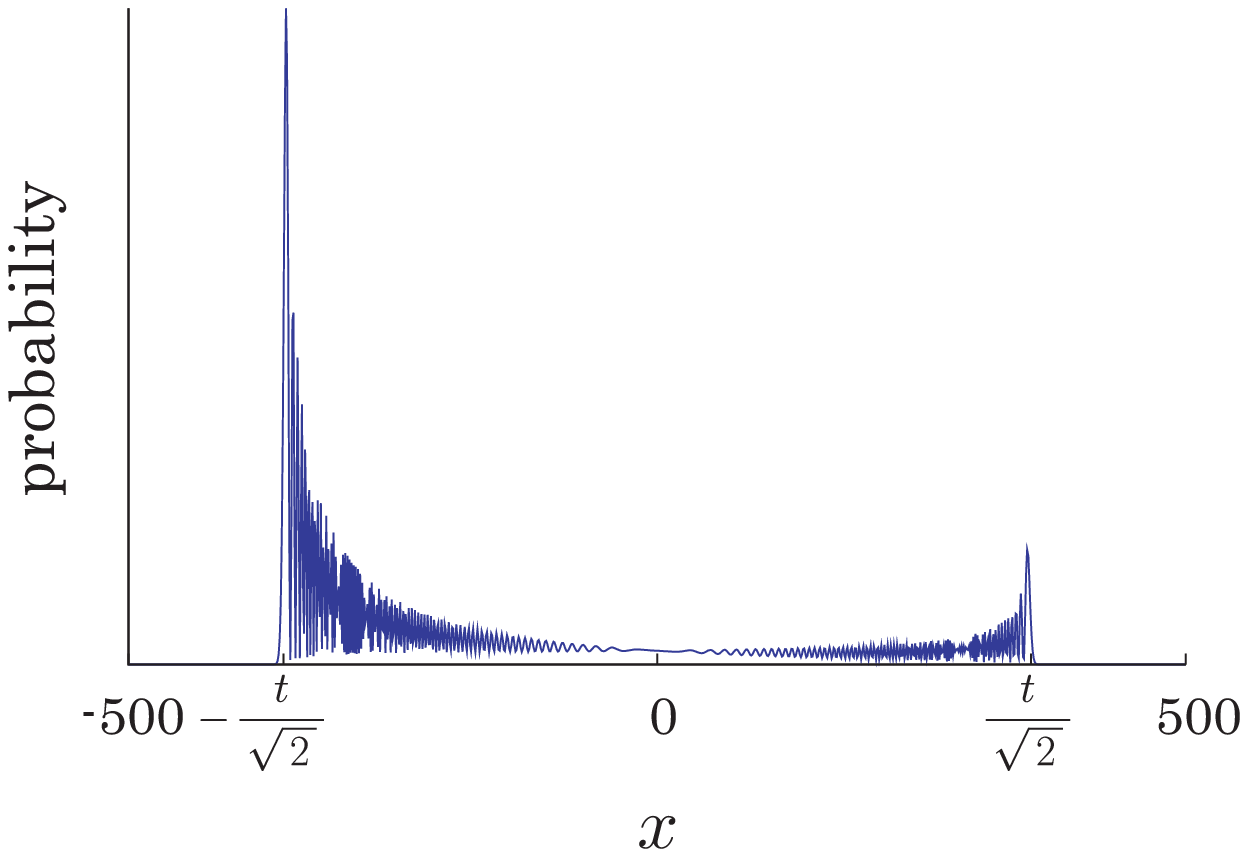}\\[2mm]
  (a)--1
  \end{center}
 \end{minipage}
 \begin{minipage}{50mm}
  \begin{center}
   \includegraphics[scale=0.4]{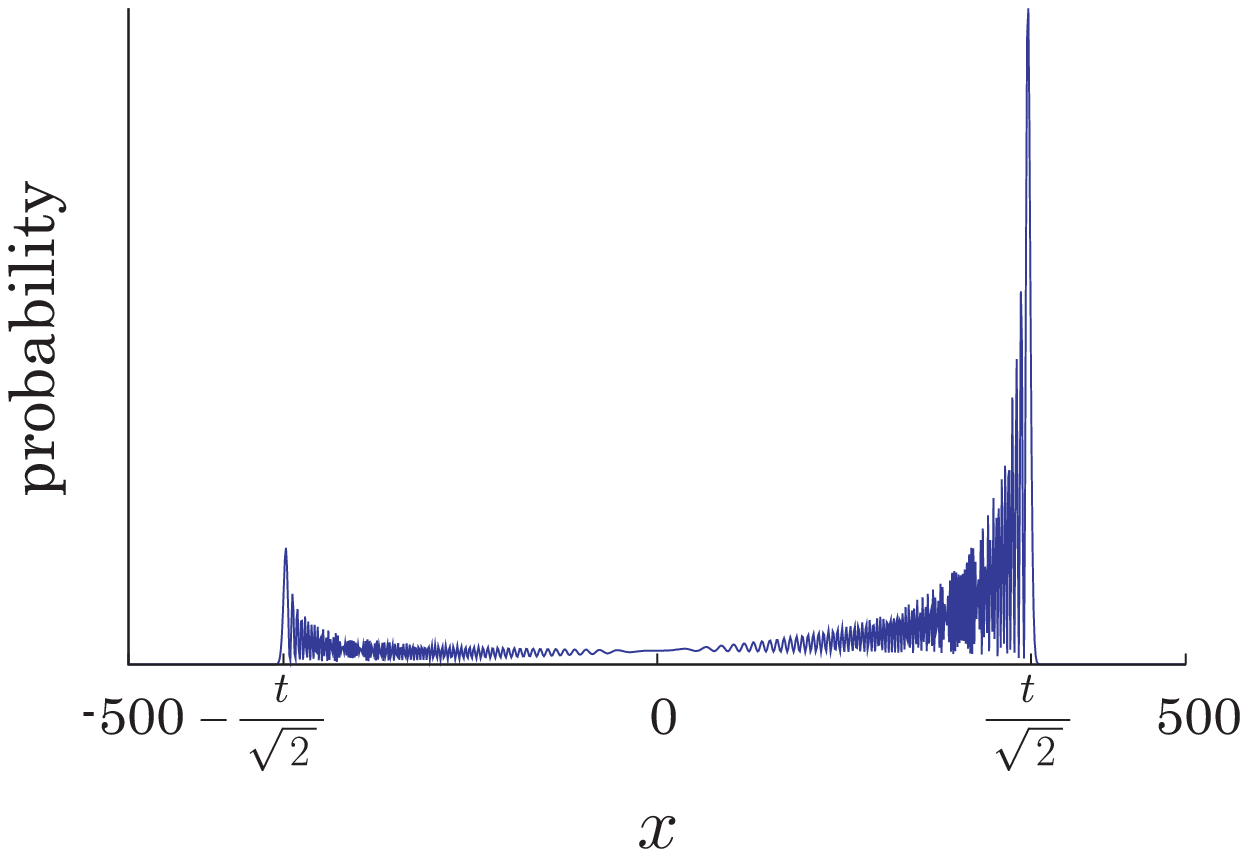}\\[2mm]
  (b)--1
  \end{center}
 \end{minipage}
 \bigskip
 \begin{minipage}{50mm}
  \begin{center}
   \includegraphics[scale=0.4]{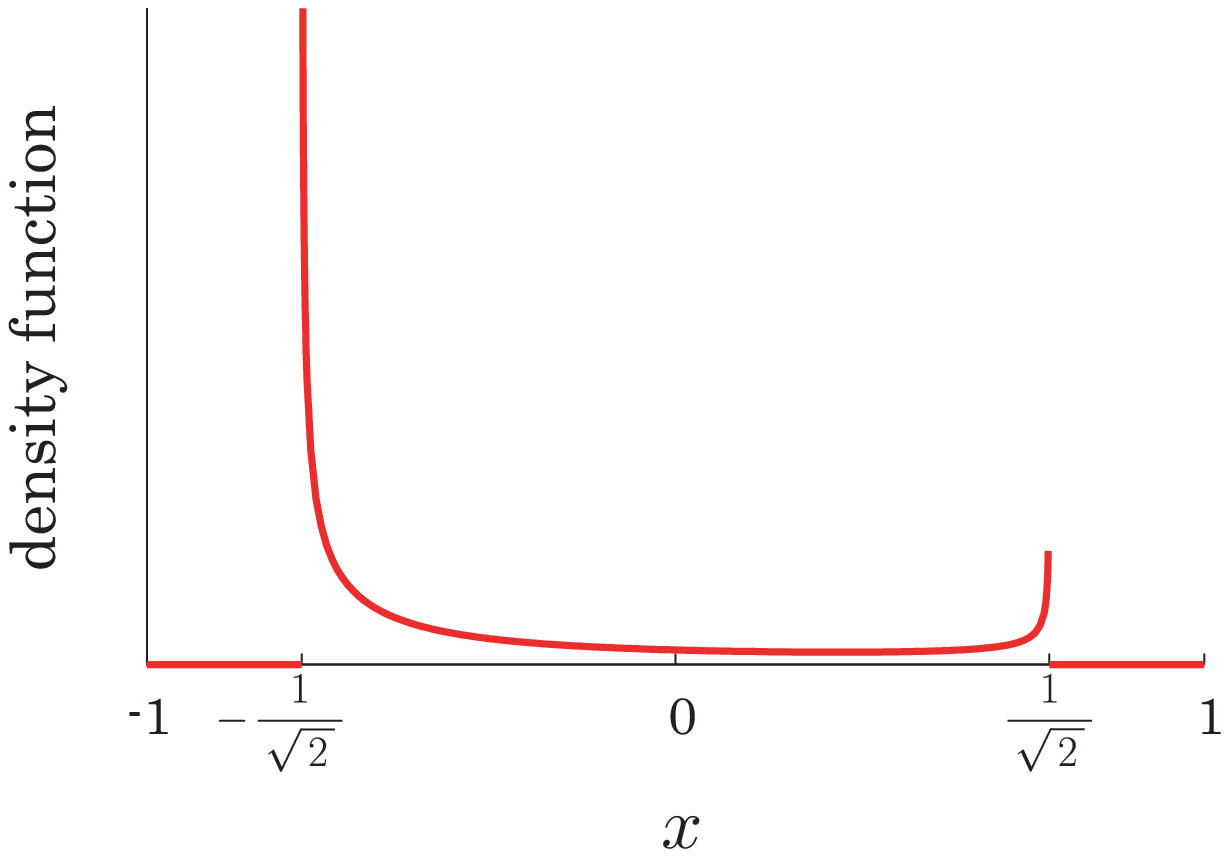}\\[2mm]
  (a)--2
  \end{center}
 \end{minipage}
 \begin{minipage}{50mm}
  \begin{center}
   \includegraphics[scale=0.4]{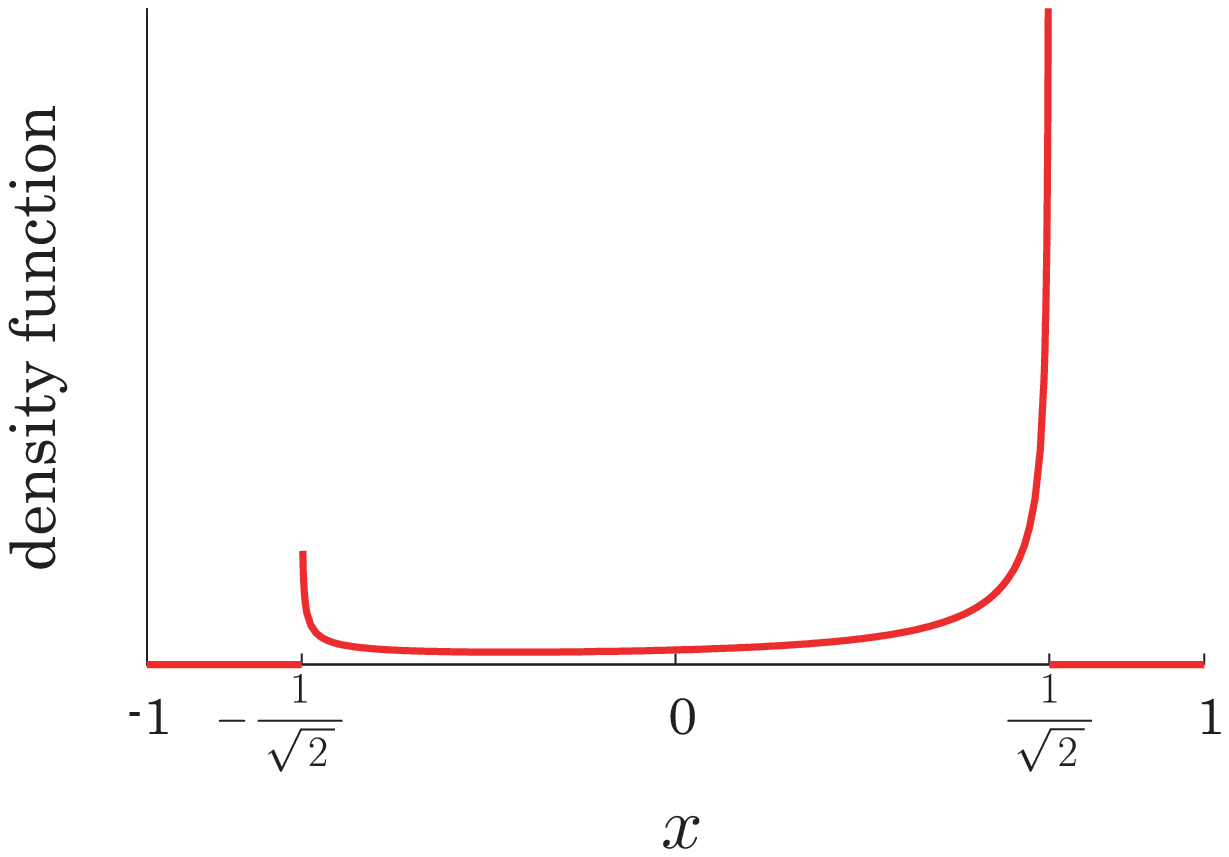}\\[2mm]
  (b)--2
  \end{center}
 \end{minipage}
 \bigskip
\begin{minipage}{50mm}
  \begin{center}
   \includegraphics[scale=0.4]{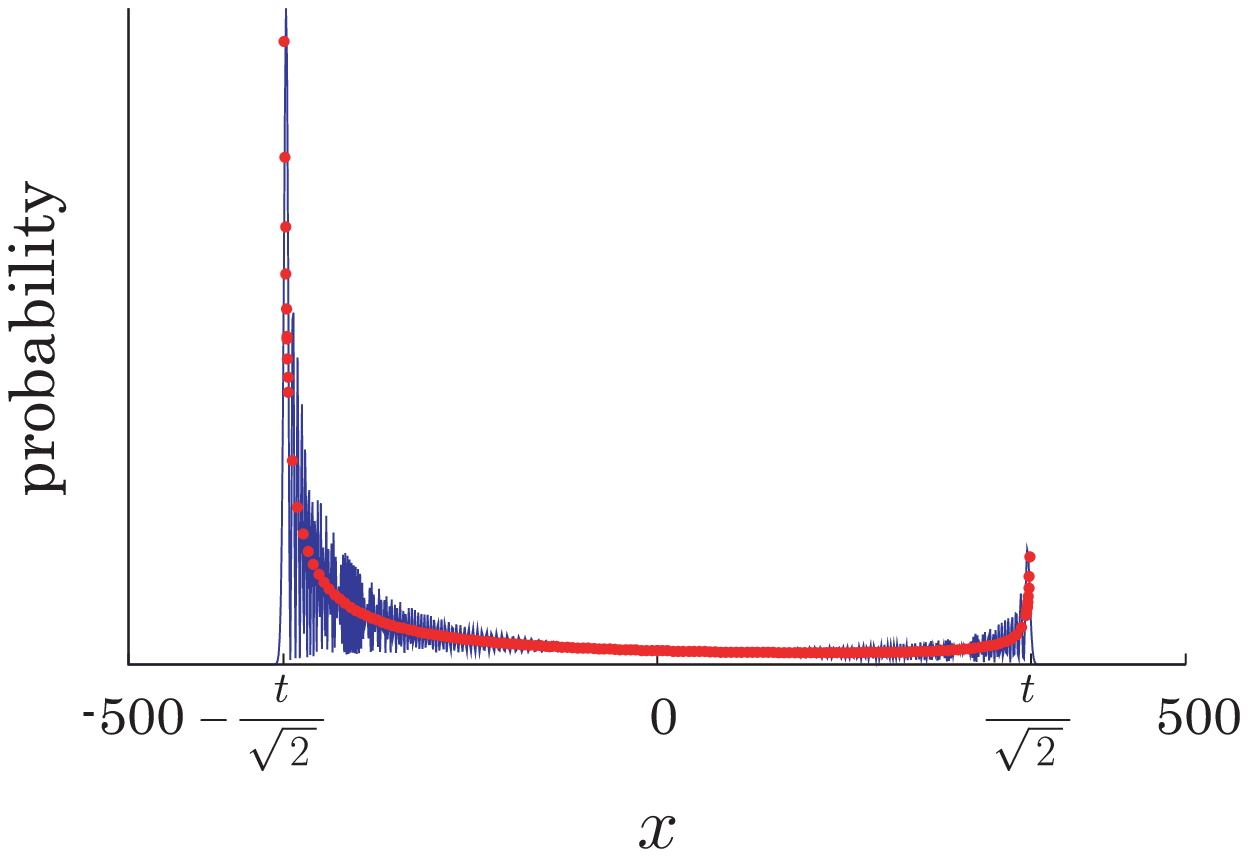}\\[2mm]
  (a)--3
  \end{center}
 \end{minipage}
 \begin{minipage}{50mm}
  \begin{center}
   \includegraphics[scale=0.4]{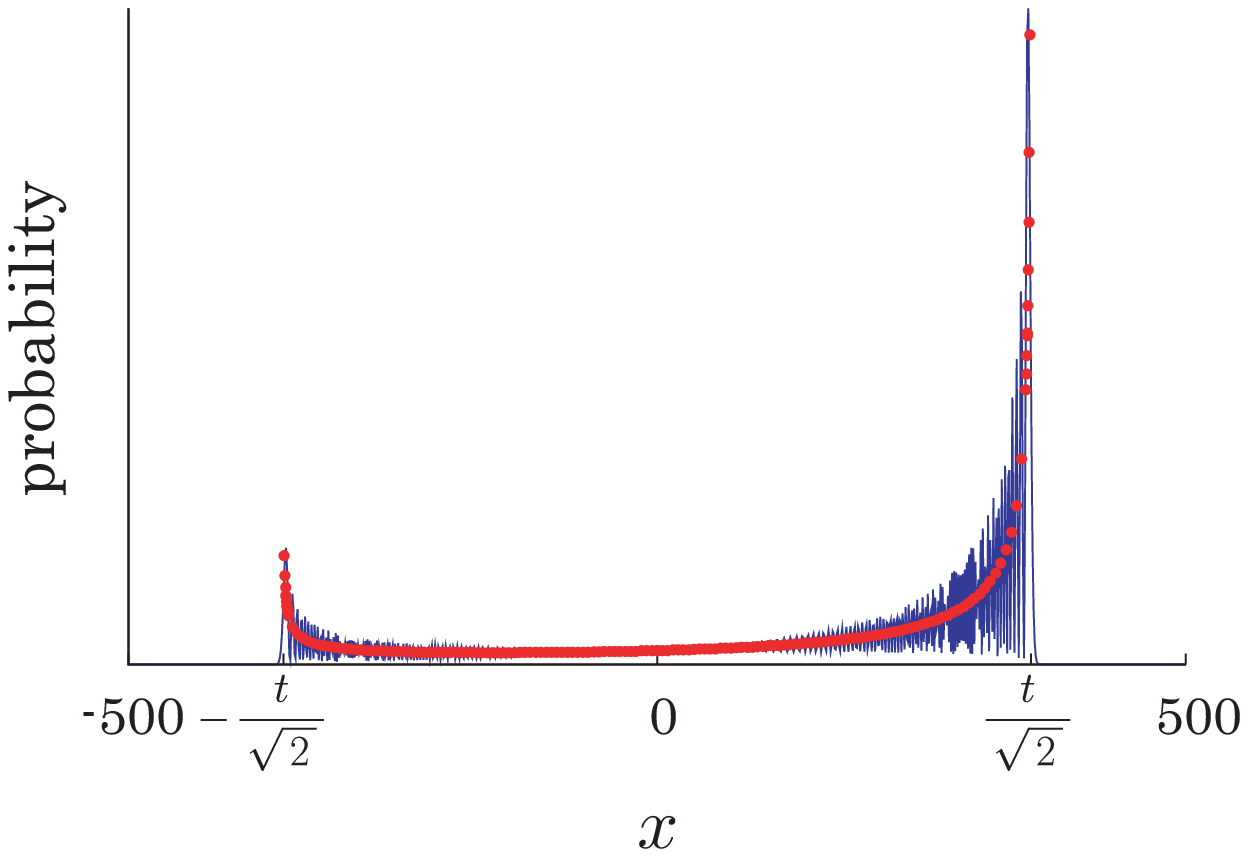}\\[2mm]
  (b)--3
  \end{center}
 \end{minipage}
\fcaption{$\rho=1/\sqrt{2}, \nu=\pi/2$ :
The blue lines represent the probability distribution $\mathbb{P}(X_t=x)$ at time $t=500$ ((a)--1, (b)--1) and the red lines represent the limit density function ((a)--2, (b)--2). 
In (a)--3 and (b)--3, we confirm that the approximation (red points) obtained from the limit density function reproduces the features of the probability distribution as time $t$ becomes large enough.
The walker launches with the localized initial state at the origin, $\ket{\Psi_0}=\ket{0}\otimes (\alpha\ket{0}+\beta\ket{1})$.}
\label{fig:1}
\end{center}
\end{figure}

This result is very similar to the usual limit theorem for a coined walk, and is contained in~\cite{Konno2008}.
The fact is also visualized in Fig.~\ref{fig:1}.
A more interesting result is obtained when the parameters take values different from the ones above.

\clearpage

\subsection{$(\rho, \nu)\neq (1/\sqrt{2}, \pi/2+n\pi)\quad(n\in\mathbb{Z})$}
\label{sec:limit_CMV}

Except for $(\rho, \nu)= (1/\sqrt{2}, \pi/2+n\pi)\quad(n\in\mathbb{Z})$, we get a more complicated formula of the limit distribution than Eq.~\eqref{eq:limit_standard_QW}.

\begin{thm}
\label{th:limit_CMV}
Assumed $(\rho, \nu)\neq (1/\sqrt{2}, \pi/2+n\pi)\,(n\in\mathbb{Z})$ and given the localized initial state of the form $\ket{\Psi_0}=\ket{0}\otimes\left(\alpha\ket{0}+\beta\ket{1}\right)$\,($\alpha,\beta\in\mathbb{C}$ such that $|\alpha|^2+|\beta|^2=1$), the quantum walker converges in distribution.
For any real number $x$, we have
 \begin{align}
  \lim_{t\to\infty}\mathbb{P}\left(\frac{X_t}{t}\leq x\right)
  =&\int_{-\infty}^x \frac{\sqrt{\eta_{+}(y)} + \sqrt{\eta_{-}(y)}}{2\pi(1-y^2)\sqrt{\xi(y)}}\,\gamma(y)\, I_{(-h^\ast, h^\ast)}(y)\,dy,
 \end{align}
 where
 \begin{align}
  \xi(x)=& (\rho^2-x^2)(\rho_0^2-x^2)-\rho^2\rho_0^2(\cos^2\nu)\, x^2,\label{eq:xi(x)}\\
  \eta_{\pm}(x)=& 1-\rho^2\rho_0^2(1+\sin^2\nu)-\left(1-\rho^2\rho_0^2\cos^2\nu\right)\,x^2\nonumber\\
  & \pm 2\rho\rho_0\,(\sin\nu)\sqrt{\xi(x)},\label{eq:eta_pm(x)}\\
  \gamma(x)=& 1+\Biggl\{\left|\alpha\right|^2-\left|\beta\right|^2-\frac{2\rho_0}{\rho}\Bigl(\Re(\alpha\overline{\beta})\cos\nu-\Im(\alpha\overline{\beta})\sin\nu\Bigr)\Biggr\}\,x,\label{eq:gamma(x)}
 \end{align}
 and
 \begin{equation}
  I_{(-h^\ast, h^\ast)}(x)=\left\{\begin{array}{cl}
	   1&(x\in (-h^\ast, h^\ast))\\
		  0&(x\notin (-h^\ast, h^\ast))
		 \end{array}\right.,\label{eq:indicator}
 \end{equation}
 with
 \begin{equation}
  h^\ast=\frac{\,1\,}{\,2\,}\Biggl\{\,\sqrt{\left(1+\rho\rho_0\,\right)^2-\rho^2\rho_0^2\sin^2\nu}-\sqrt{\left(1-\rho\rho_0\,\right)^2-\rho^2\rho_0^2\sin^2\nu}\,\Biggr\}.
 \end{equation}
The notation $\Re(z)$ denotes the real part of a complex number $z$.
\end{thm}

\bigskip

The proof of this theorem is demonstrated in Appendix.
Looking at Fig.~\ref{fig:2}, we realize the difference from Fig.~\ref{fig:1} which was for the case of $(\rho, \nu)= (1/\sqrt{2}, \pi/2+n\pi)\,(n\in\mathbb{Z})$.
The probability distributions in the pictures (a)--1 and (b)--1 vibrate harder than the ones of the standard coined walk in Fig.~\ref{fig:1}.

\clearpage

\begin{figure}[h]
\begin{center}
\begin{minipage}{50mm}
  \begin{center}
  (a) $(\alpha, \beta) = (1/\sqrt{2}\,,i/\sqrt{2}\,)$
  \end{center}
 \end{minipage}
 \begin{minipage}{50mm}
  \begin{center}
  (b) $(\alpha, \beta) = (1,0)$
  \end{center}
 \end{minipage}\\
 \bigskip
 \begin{minipage}{50mm}
  \begin{center}
   \includegraphics[scale=0.4]{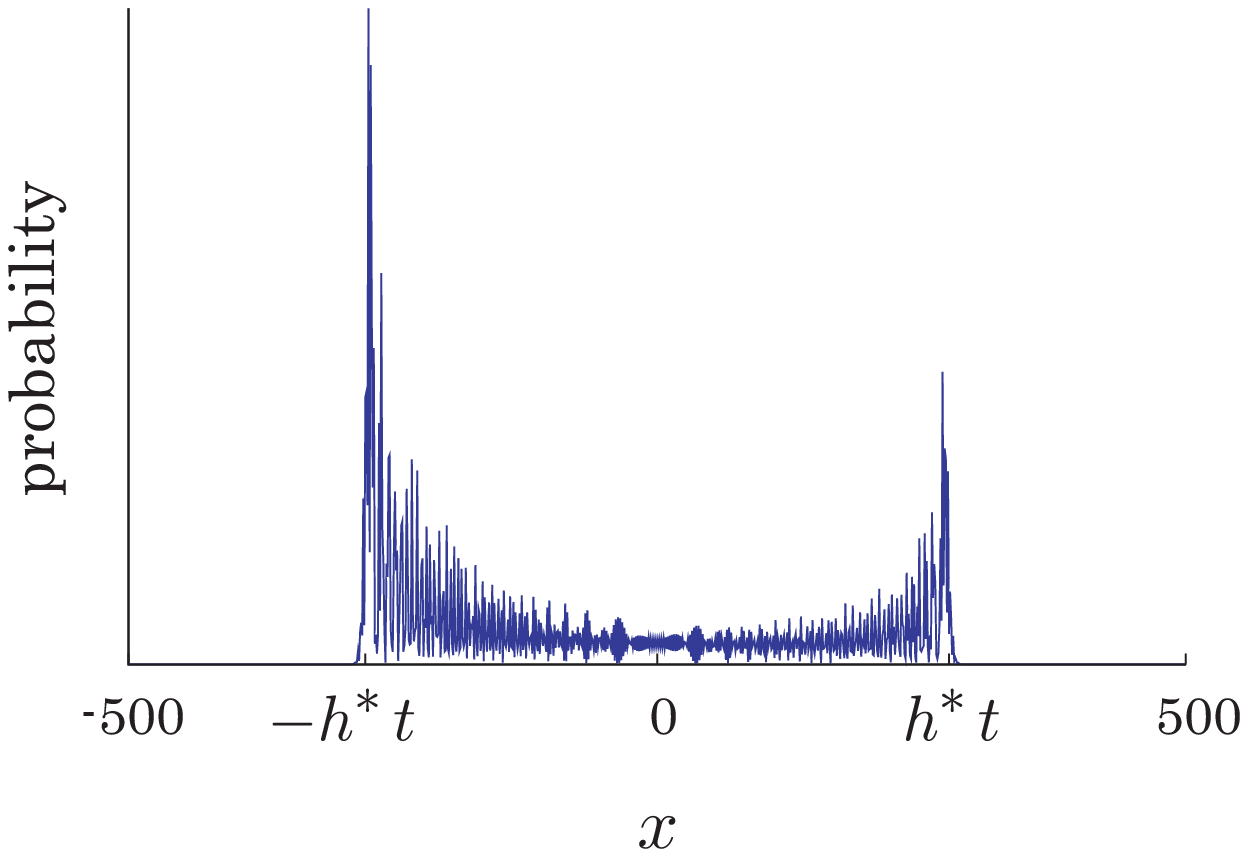}\\[2mm]
  (a)--1
  \end{center}
 \end{minipage}
 \begin{minipage}{50mm}
  \begin{center}
   \includegraphics[scale=0.4]{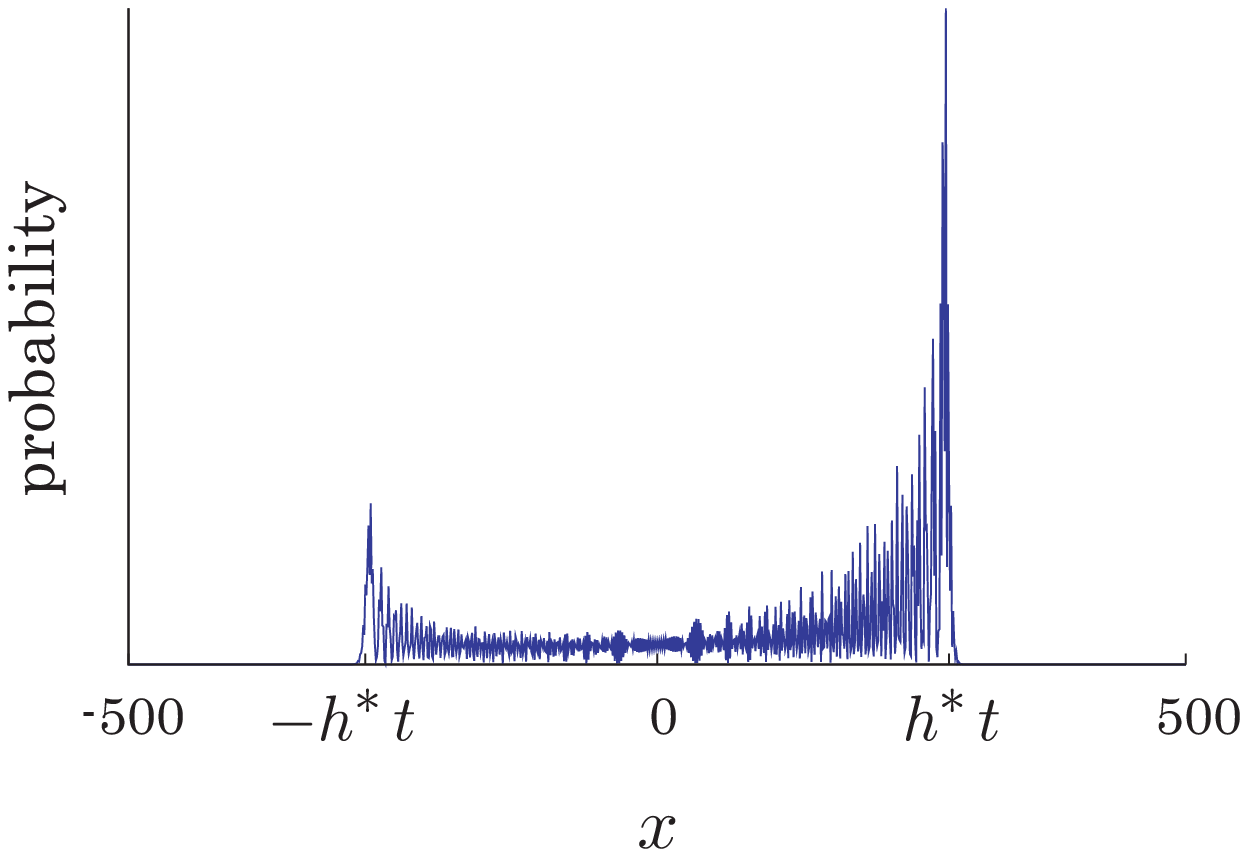}\\[2mm]
  (b)--1
  \end{center}
 \end{minipage}
 \bigskip
 \begin{minipage}{50mm}
  \begin{center}
   \includegraphics[scale=0.4]{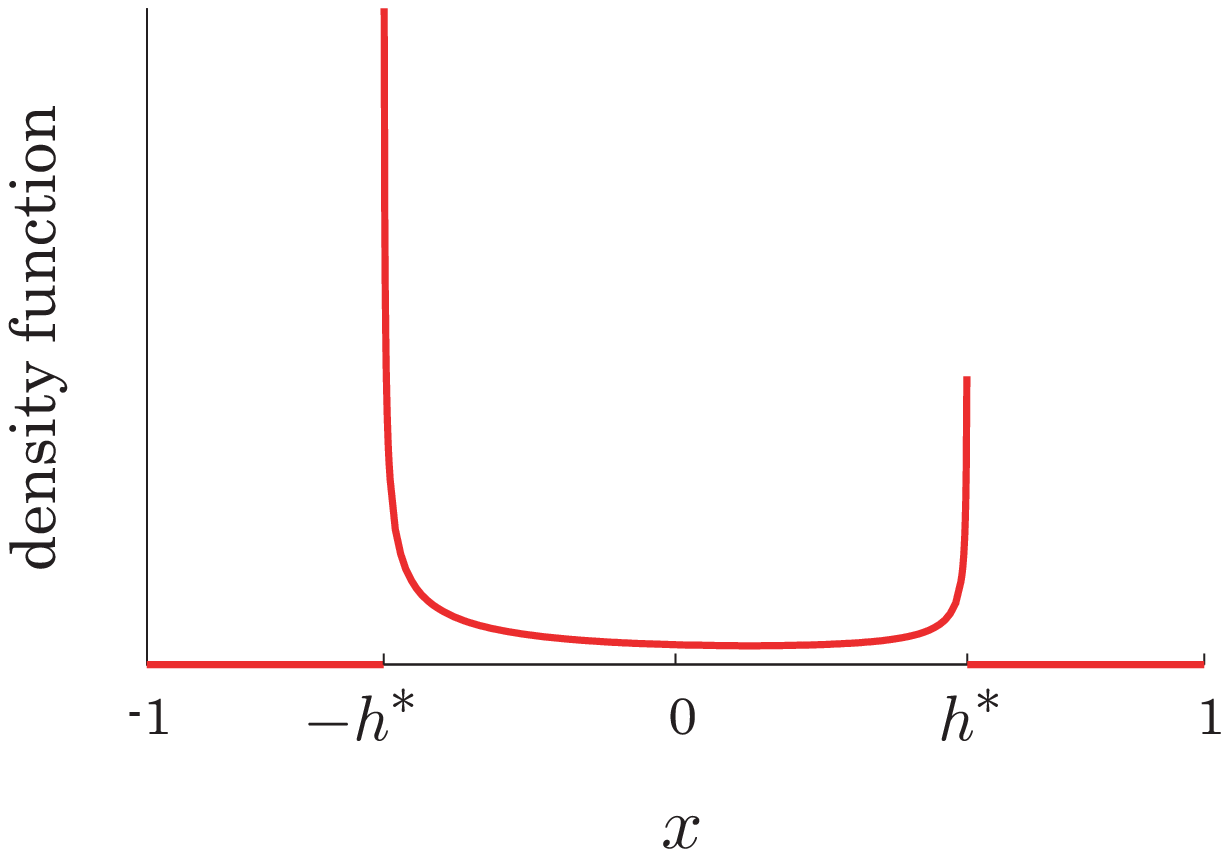}\\[2mm]
  (a)--2
  \end{center}
 \end{minipage}
 \begin{minipage}{50mm}
  \begin{center}
   \includegraphics[scale=0.4]{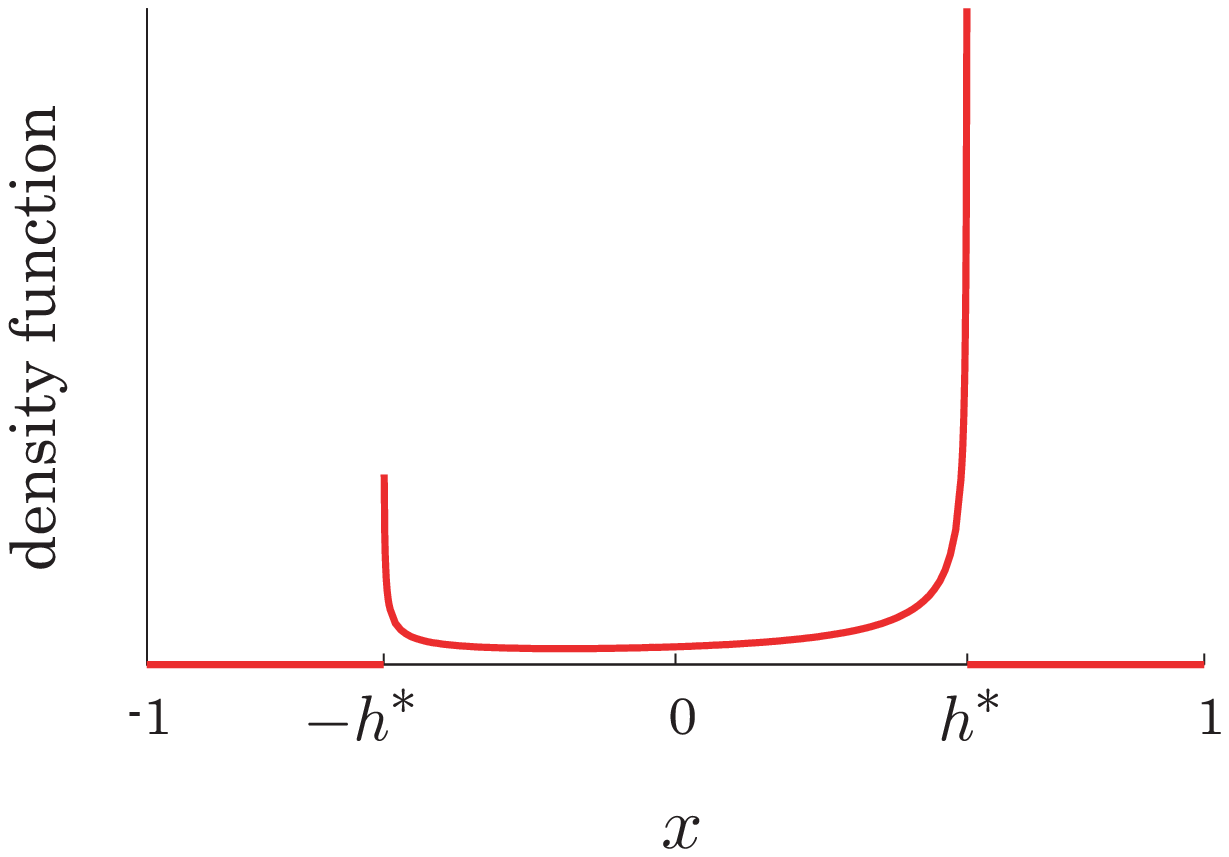}\\[2mm]
  (b)--2
  \end{center}
 \end{minipage}
 \bigskip
\begin{minipage}{50mm}
  \begin{center}
   \includegraphics[scale=0.4]{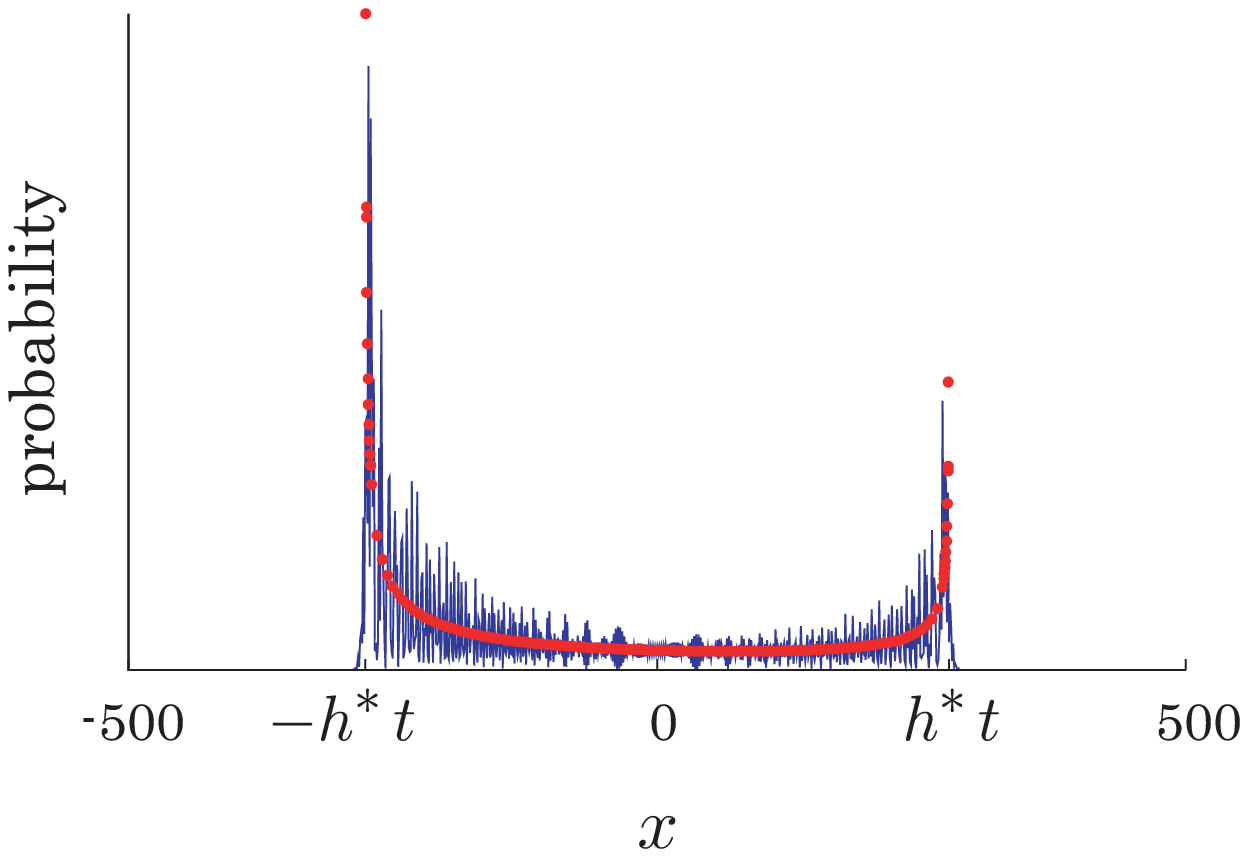}\\[2mm]
  (a)--3
  \end{center}
 \end{minipage}
 \begin{minipage}{50mm}
  \begin{center}
   \includegraphics[scale=0.4]{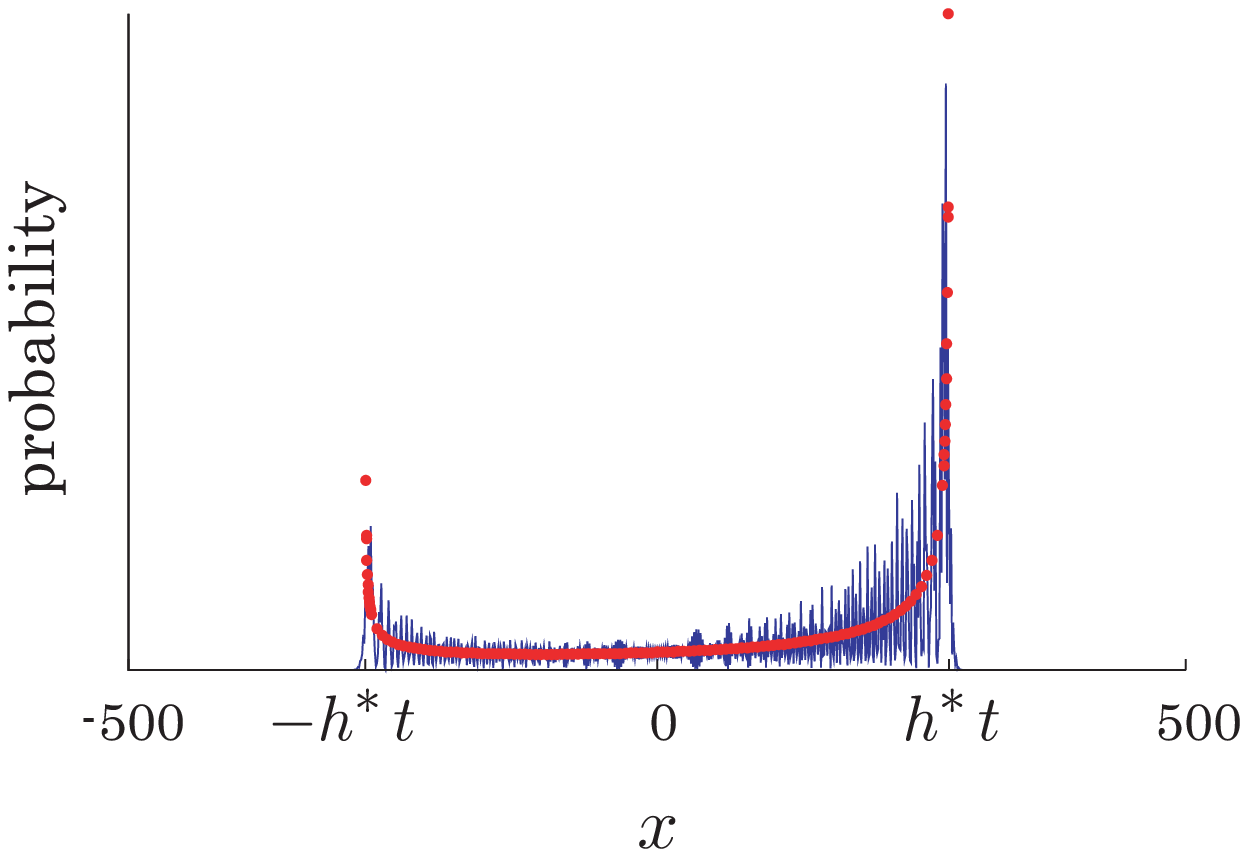}\\[2mm]
  (b)--3
  \end{center}
 \end{minipage}
\fcaption{$\rho=1/\sqrt{2}, \nu=\pi/4$ :
The blue lines represent the probability distribution $\mathbb{P}(X_t=x)$ at time $t=500$ ((a)--1, (b)--1) and the red lines represent the limit density function ((a)--2, (b)--2). 
In (a)--3 and (b)--3, we confirm that the approximation (red points) obtained from the limit density function reproduces the features of the probability distribution as time $t$ becomes large enough.
The walker launches with the localized initial state at the origin, $\ket{\Psi_0}=\ket{0}\otimes (\alpha\ket{0}+\beta\ket{1})$.}
\label{fig:2}
\end{center}
\end{figure}

\clearpage

\begin{figure}[h]
\begin{center}
  \begin{minipage}[b]{60mm}
  \begin{center}
   \includegraphics[scale=0.5]{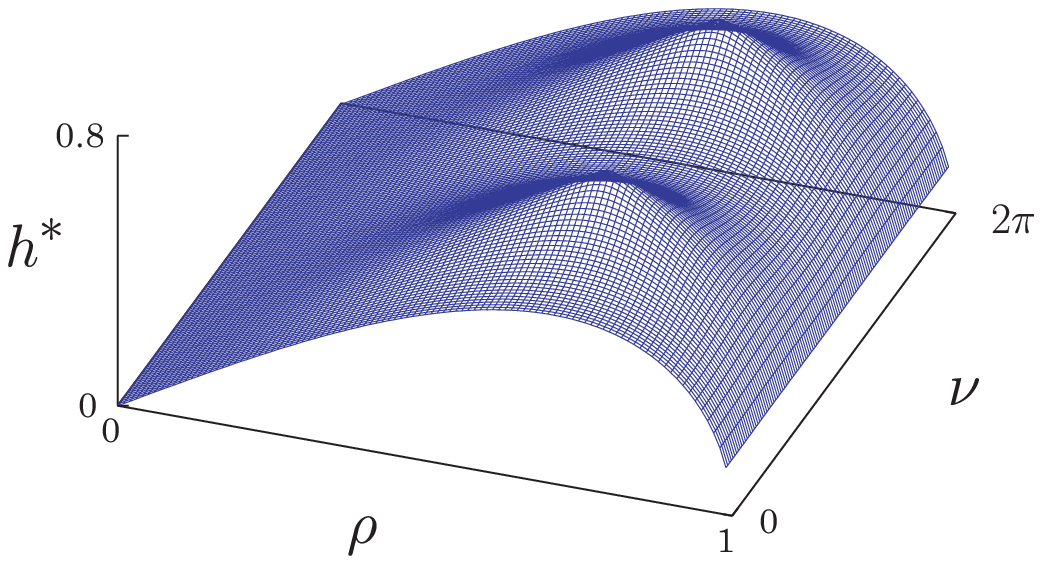}\\[10mm]
  (a)
  \end{center}
 \end{minipage}
 \begin{minipage}[b]{50mm}
  \begin{center}
   \includegraphics[scale=0.3]{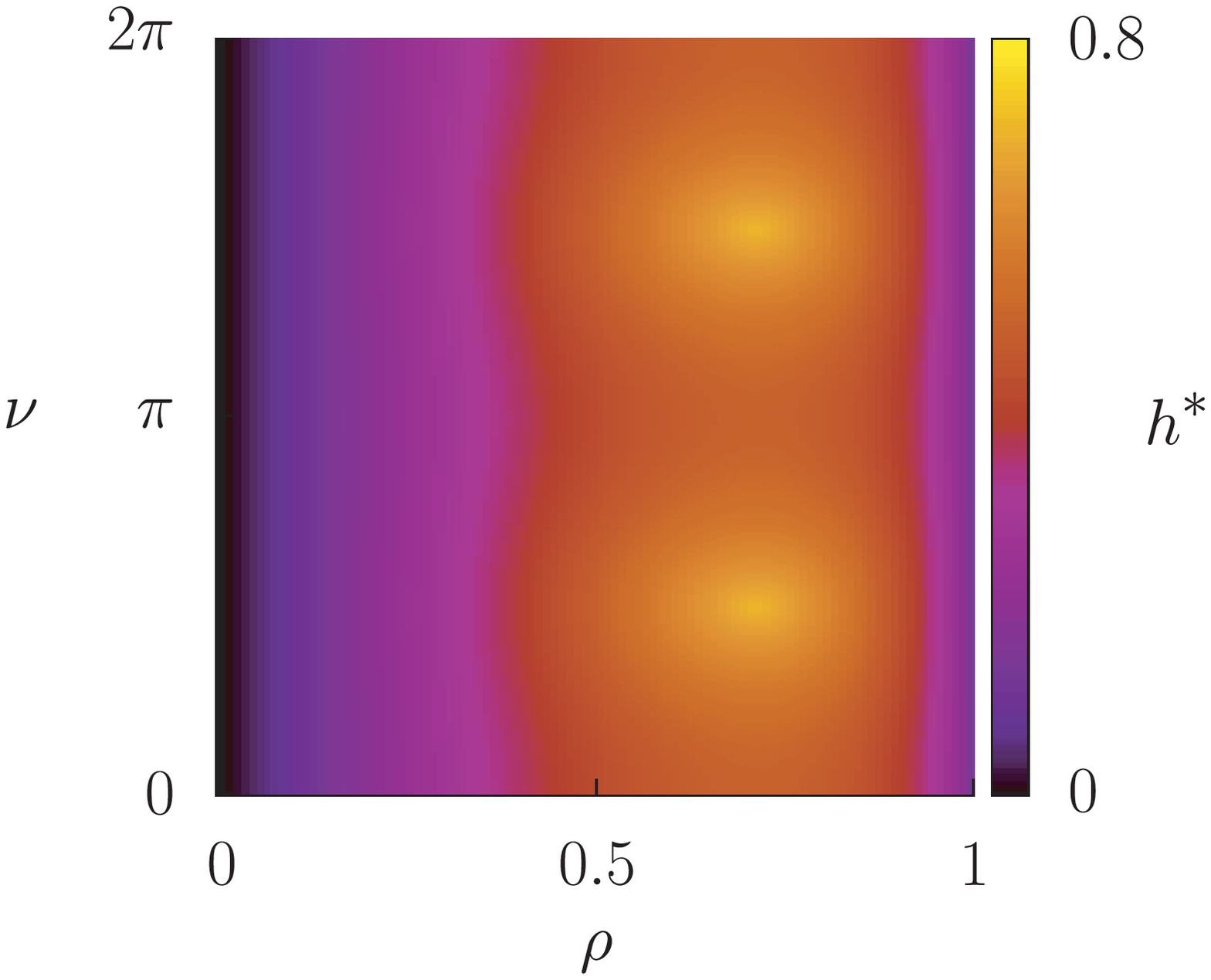}\\[2mm]
  (b)
  \end{center}
 \end{minipage}
\fcaption{The support of the limit density function depends on the parameters $\rho$ and $\nu$ which determine the five-diagonal matrix. The picture (b) is another expression of the picture (a) in 2D density plot.}
\label{fig:4}
\end{center}
\end{figure}

Moreover, while the limit density function of the standard coined walk has a support and the support does not depend on the phase term of the operation~\cite{Konno2008}, the phase term $\nu$ gives an effect on the quantity $h^\ast$ which defines the support of the limit density function, as Fig.~\ref{fig:4} shows.

\bigskip

Related to a past study, the limit distribution is consistent with Eq.~(16) in Machida and Konno~\cite{MachidaKonno2010} if we fix the value of $\nu$ at $0$ or $\pi$.
The quantum walk is a two-period time-dependent coined quantum walk alternately driven by two orthogonal matrices in~\cite{MachidaKonno2010},
\begin{equation}
 H_0=\begin{bmatrix}
      \rho & -\rho_0\\
      -\rho_0 & -\rho 
     \end{bmatrix},\quad
  H_1=\begin{bmatrix}
   -\rho_0 & \rho\\
   -\rho & -\rho_0 
  \end{bmatrix}.
\end{equation}
Note that the form of the limit density function presented in this paper is a different form from the one in~\cite{MachidaKonno2010} due to the phase parameter $\nu$, and it has not been mathematically discovered in quantum walks before.  

Since the substitution $\rho=1/\sqrt{2}$ and $\nu=\pi/2+n\pi\,(n\in\mathbb{Z})$ makes
the functions $\xi(x)=(1-2x^2)^2/4$, $\sqrt{\eta_{+}(x)}+\sqrt{\eta_{-}(x)}=\sqrt{1-2x^2}$, $\gamma(x)=1+\bigl\{\left|\alpha\right|^2-\left|\beta\right|^2+(-1)^n \cdot 2\Im(\alpha\overline{\beta})\bigr\}x=\Theta_n(x)$, and $h^\ast=1/\sqrt{2}$, we fulfill Eq.~\eqref{eq:limit_standard_QW} again, which means that the limit distribution for the case $(\rho, \nu)= (1/\sqrt{2}, \pi/2+n\pi)\,(n\in\mathbb{Z})$ is allowed to be combined into Theorem~\ref{th:limit_CMV}.


\section{Summary}
We studied a quantum walk on the line which evolved with a matrix, and demonstrated long-time limit distributions.
The quantum walker did not localize at all and spread away as time $t$ became large.
The limit distributions, moreover, approximately depicted the probability distributions of the quantum walk, shown as Figs.~\ref{fig:2} and \ref{fig:3}.
While some special values of parameters $\theta$ and $\nu$ produced a standard quantum walk whose limit distributions were already derived in~\cite{Konno2008}, the others gave a different type of limit distribution as we can expect from the bigger oscillation in probability distributions of the walk in Fig~\ref{fig:2}.
The study in this paper told us an explicit form of the limit distribution for the quantum walk.
\bigskip

\begin{center}
{\bf Acknowledgements}
\end{center}
The author is supported by JSPS Grant-in-Aid for Scientific Research (C) (No.19K03625) and thanks F. Alberto Gr\"{u}nbaum for useful comments on this paper.
\bigskip

\clearpage

\appendix{~The proof of Theorem~\ref{th:limit_CMV}}
\label{app:proof}

We demonstrate the proof of Theorem~\ref{th:limit_CMV}, that is, convergence of $X_t/t$ in distribution as $t\to\infty$.
The unitary matrix $\tilde{H}(k)$ holds the eigenvalues
\begin{equation}
 \lambda_j(k)=i\,\rho\rho_0(\sin k-\sin\nu)-(-1)^j\sqrt{1-\rho^2\rho_0^2(\sin k-\sin\nu)^2}\quad (j=1,2),
\end{equation}
and the eigenvalue $\lambda_j(k)$ is associated with the eigenvectors in a normalized expression
\begin{equation}
 \ket{v_j(k)}
  =\frac{1}{\sqrt{N_j(k)}}
  \begin{bmatrix}
   -\rho^2 e^{i\nu}+\rho_0^2e^{-ik}\\
   \rho\rho_0(e^{i\nu}+e^{-ik})-(-1)^j\sqrt{J(k)}
  \end{bmatrix},
\end{equation}
where
\begin{align}
 N_j(k)=&2\left\{J(k)-(-1)^j \rho\rho_0(\cos k+\cos\nu)\sqrt{J(k)}\right\},\\
 J(k)=&1-\rho^2\rho_0^2(\sin k-\sin\nu)^2.
\end{align}
Since $J(k)$ is larger than $0$ for any $k\in [-\pi,\pi)$ as long as $(\rho, \nu)\neq (1/\sqrt{2}, \pi/2+n\pi)\,(n\in\mathbb{Z})$, the normalizing factors $N_j(k)\,(j=1,2)$ keep to be larger than $0$.

Once the initial state is expanded as $\ket{\tilde\phi}=\sum_{j=1}^2 \braket{v_j(k)|\tilde\phi}\ket{v_j(k)}$ in the eigenspace of matrix $\tilde{H}(k)$, the Fourier transform gets the expression in the eigenspace,
\begin{equation}
 \ket{\hat\psi_t(k)}=\sum_{j=1}^2 \lambda_j(k)^t\braket{v_j(k)|\tilde\phi}\ket{v_j(k)}.
\end{equation}
With $D=i\cdot d/dk$ and the Pochhammer notation $(t)_r=t\cdot (t-1)\times\cdots\times(t-r+1)$, one approaches the $r$-th moments ($r=0,1,2,\ldots$)
\begin{align}
 \mathbb{E}[X_t^r]
 =&\int_{-\pi}^{\pi} \bra{\hat\psi_t(k)} \Bigl(D^r\ket{\hat\psi_t(k)}\Bigr)\frac{dk}{2\pi}\nonumber\\
 =&(t)_r\left\{\sum_{j=1}^2 \int_{-\pi}^{\pi} \left(\frac{i\,\lambda'_j(k)}{\lambda_j(k)}\right)^r \Bigl|\braket{v_j(k)|\tilde\phi}\Bigr|^2 \frac{dk}{2\pi}\right\}+O(t^{r-1}),\nonumber
\end{align}
from which
\begin{equation}
 \lim_{t\to\infty}\mathbb{E}\left[\left(\frac{X_t}{t}\right)^r\right]
  =\sum_{j=1}^2 \int_{-\pi}^{\pi} \left(\frac{i\,\lambda'_j(k)}{\lambda_j(k)}\right)^r \Bigl|\braket{v_j(k)|\tilde\phi}\Bigr|^2 \frac{dk}{2\pi},\label{eq:moment_converge}
\end{equation}
where the functions $i\,\lambda'_j(k)/\lambda_j(k)$ are computed to be of the form
\begin{equation}
 \frac{i\,\lambda'_j(k)}{\lambda_j(k)}=(-1)^j\frac{\rho\rho_0\cos k}{\sqrt{1-\rho^2\rho_0^2(\sin k-\sin \nu)^2}}\quad(j=1,2).
\end{equation}
Putting $h(\nu;k)=\rho\rho_0\cos k/\sqrt{1-\rho^2\rho_0^2(\sin k-\sin \nu)^2}$, we are going to transform Eq.~\eqref{eq:moment_converge}.
First, the range of the function $h(k)$ is computed to be $[-h^\ast,\,h^\ast]$ where
\begin{align}
 h^\ast=&h(\nu;k^\ast)=\frac{\,1\,}{\,2\,}\Biggl\{\,\sqrt{\left(1+\rho\rho_0\,\right)^2-\rho^2\rho_0^2\sin^2\nu}-\sqrt{\left(1-\rho\rho_0\,\right)^2-\rho^2\rho_0^2\sin^2\nu}\,\Biggr\},\\
 k^\ast=&\left\{\begin{array}{ll}
	  \arcsin\Biggl[\frac{-1+\rho^2\rho_0^2(1+\sin^2\nu)+\sqrt{\left\{(1-\rho\rho_0)^2-\rho^2\rho_0^2\sin^2\nu\right\}\left\{(1+\rho\rho_0)^2-\rho^2\rho_0^2\sin^2\nu\right\}}}{2\rho^2\rho_0^2\sin\nu}\Biggr] & (\sin\nu\neq 0)\\[5mm]
		 0 & (\sin\nu=0)
		\end{array}\right..
\end{align}
Then, we shrink the range of the integrals
\begin{align}
  \int_{-\pi}^{\pi} \left(\frac{i\,\lambda'_j(k)}{\lambda_j(k)}\right)^r \Bigl|\braket{v_j(k)|\tilde\phi}\Bigr|^2 \frac{dk}{2\pi}
 =& \int_{-\pi}^{\pi} \left\{(-1)^j h(\nu;k)\right\}^r \Bigl|\braket{v_j(k)|\tilde\phi}\Bigr|^2 \frac{dk}{2\pi}\nonumber\\
 =& \int_{-\pi}^{-\pi/2} \quad + \quad \int_{-\pi/2}^{\pi/2} \quad + \quad \int_{\pi/2}^{\pi}\nonumber\\
 =& \int_{\pi}^{3\pi/2} \quad + \quad \int_{-\pi/2}^{\pi/2} \quad + \quad \int_{\pi/2}^{\pi}\nonumber\\
 =& \int_{-\pi/2}^{\pi/2} \quad + \quad \int_{\pi/2}^{3\pi/2}\nonumber\\
 =& \int_{-\pi/2}^{\pi/2} \left\{(-1)^j h(\nu;k)\right\}^r \Bigl|\braket{v_j(k)|\tilde\phi}\Bigr|^2 \frac{dk}{2\pi}\nonumber\\
  &+ \int_{-\pi/2}^{\pi/2} \left\{(-1)^j h(\nu;\pi-k)\right\}^r \Bigl|\braket{v_j(\pi-k)|\tilde\phi}\Bigr|^2 \frac{dk}{2\pi}\nonumber\\
 =& \int_{-\pi/2}^{\pi/2} \left\{(-1)^j h(\nu;k)\right\}^r \Bigl|\braket{v_j(k)|\tilde\phi}\Bigr|^2 \frac{dk}{2\pi}\nonumber\\
  &+ \int_{-\pi/2}^{\pi/2} \left\{-(-1)^j h(\pi-\nu;k)\right\}^r \Bigl|\braket{v_j(\pi-k)|\tilde\phi}\Bigr|^2 \frac{dk}{2\pi}\nonumber\\
 =& \int_{-\pi/2}^{\pi/2} \left\{(-1)^j h(\nu;k)\right\}^r \Bigl|\braket{v_j(k)|\tilde\phi}\Bigr|^2 \frac{dk}{2\pi}\nonumber\\
  &+ \int_{-\pi/2}^{\pi/2} \left\{-(-1)^j h(\nu;k)\right\}^r \Bigl|\braket{v_j(\pi-k)|\tilde\phi}\Bigr|^2 \frac{dk}{2\pi}.
\end{align}
\clearpage
Introducing the functions
\begin{align}
 J(\nu;k) =& 1-\rho^2\rho_0^2(\sin k-\sin\nu)^2,\\
 F_{0,\pm}(\nu;k) =& \frac{1}{2}\mp \frac{\rho\rho_0(\cos k+\cos\nu)}{2\sqrt{J(\nu;k)}},\\
 F_{1,\pm}(\nu;k) =& \frac{1}{2}\pm \frac{\rho\rho_0(\cos k+\cos\nu)}{2\sqrt{J(\nu;k)}},\\
 F_{2,\pm}(\nu;k) =& \pm \frac{\rho_0^2\cos k -\rho^2\cos\nu}{\sqrt{J(\nu;k)}},\\
 F_{3,\pm}(\nu;k) =& \pm \frac{\rho_0^2\sin k +\rho^2\sin\nu}{\sqrt{J(\nu;k)}},
\end{align}
we compute the integrals more concretely,
\begin{align}
 & \sum_{j=1}^2 \int_{-\pi}^{\pi} \left(\frac{i\,\lambda'_j(k)}{\lambda_j(k)}\right)^r \Bigl|\braket{v_j(k)|\tilde\phi}\Bigr|^2 \frac{dk}{2\pi}\nonumber\\
 = & \int_{-\pi/2}^{\pi/2} h(\nu;k)^r\Bigl[\left\{F_{0,-}(\nu;k)+F_{0,-}(\pi-\nu;k)\right\}|\alpha|^2\nonumber\\
 & \hspace{24mm} +\left\{F_{1,-}(\nu;k)+F_{1,-}(\pi-\nu;k)\right\}|\beta|^2\nonumber\\
 & \hspace{24mm} +\left\{F_{2,-}(\nu;k)+F_{2,-}(\pi-\nu;k)\right\}\Re(\alpha\overline{\beta}e^{i\nu})\nonumber\\
 & \hspace{24mm} -\left\{F_{3,-}(\nu;k)-F_{3,-}(\pi-\nu;k)\right\}\Im(\alpha\overline{\beta}e^{i\nu})\Bigr]\,\frac{dk}{2\pi}\nonumber\\
 & + \int_{-\pi/2}^{\pi/2} (-h(\nu;k))^r\Bigl[\left\{F_{0,+}(\nu;k)+F_{0,+}(\pi-\nu;k)\right\}|\alpha|^2\nonumber\\
 & \hspace{32mm} +\left\{F_{1,+}(\nu;k)+F_{1,+}(\pi-\nu;k)\right\}|\beta|^2\nonumber\\
 & \hspace{32mm} +\left\{F_{2,+}(\nu;k)+F_{2,+}(\pi-\nu;k)\right\}\Re(\alpha\overline{\beta}e^{i\nu})\nonumber\\
 & \hspace{32mm} -\left\{F_{3,+}(\nu;k)-F_{3,+}(\pi-\nu;k)\right\}\Im(\alpha\overline{\beta}e^{i\nu})\Bigr]\,\frac{dk}{2\pi},\nonumber\\
 = & \int_{-\pi/2}^{\pi/2} h(\nu;k)^r\,\biggl\{\left(1+h(\nu;k)\right)|\alpha|^2
 +\left(1-h(\nu;k)\right)|\beta|^2
 -\frac{2\rho_0}{\rho}\,h(\nu;k)\Re(\alpha\overline{\beta}e^{i\nu})\biggr\}\,\frac{dk}{2\pi}\nonumber\\
 & + \int_{-\pi/2}^{\pi/2} (-h(\nu;k))^r\,\biggl\{\left(1-h(\nu;k)\right)|\alpha|^2
 +\left(1+h(\nu;k)\right)|\beta|^2
 +\frac{2\rho_0}{\rho}\,h(\nu;k)\Re(\alpha\overline{\beta}e^{i\nu})\biggr\}\,\frac{dk}{2\pi}.
\end{align}
Changing the variable in the integrals from $k$ to $x$ by putting $h(k)=x$, we have
\begin{align}
 & \sum_{j=1}^2 \int_{-\pi}^{\pi} \left(\frac{i\,\lambda'_j(k)}{\lambda_j(k)}\right)^r \Bigl|\braket{v_j(k)|\tilde\phi}\Bigr|^2 \frac{dk}{2\pi}\nonumber\\
 = & \int_{-\pi/2}^{\pi/2} h(\nu;k)^r\,\biggl\{\left(1+h(\nu;k)\right)|\alpha|^2
 +\left(1-h(\nu;k)\right)|\beta|^2
 -\frac{2\rho_0}{\rho}\,h(\nu;k)\Re(\alpha\overline{\beta}e^{i\nu})\biggr\}\,\frac{dk}{2\pi}\nonumber\\
 & + \int_{-\pi/2}^{\pi/2} (-h(\nu;k))^r\,\biggl\{\left(1-h(\nu;k)\right)|\alpha|^2
 +\left(1+h(\nu;k)\right)|\beta|^2
 +\frac{2\rho_0}{\rho}\,h(\nu;k)\Re(\alpha\overline{\beta}e^{i\nu})\biggr\}\,\frac{dk}{2\pi}\nonumber\\
 = & \int_{-\pi/2}^{k^\ast}h(\nu;k)^r\,\biggl\{\left(1+h(\nu;k)\right)|\alpha|^2
 +\left(1-h(\nu;k)\right)|\beta|^2
 -\frac{2\rho_0}{\rho}\,h(\nu;k)\Re(\alpha\overline{\beta}e^{i\nu})\biggr\}\,\frac{dk}{2\pi}\nonumber\\
 & +\int_{k^\ast}^{\pi/2}h(\nu;k)^r\,\biggl\{\left(1+h(\nu;k)\right)|\alpha|^2
 +\left(1-h(\nu;k)\right)|\beta|^2
 -\frac{2\rho_0}{\rho}\,h(\nu;k)\Re(\alpha\overline{\beta}e^{i\nu})\biggr\}\,\frac{dk}{2\pi}\nonumber\\
 & + \int_{-\pi/2}^{k^\ast} (-h(\nu;k))^r\,\biggl\{\left(1-h(\nu;k)\right)|\alpha|^2
 +\left(1+h(\nu;k)\right)|\beta|^2
 +\frac{2\rho_0}{\rho}\,h(\nu;k)\Re(\alpha\overline{\beta}e^{i\nu})\biggr\}\,\frac{dk}{2\pi}\nonumber\\
 & + \int_{k^\ast}^{\pi/2} (-h(\nu;k))^r\,\biggl\{\left(1-h(\nu;k)\right)|\alpha|^2
 +\left(1+h(\nu;k)\right)|\beta|^2
 +\frac{2\rho_0}{\rho}\,h(\nu;k)\Re(\alpha\overline{\beta}e^{i\nu})\biggr\}\,\frac{dk}{2\pi}\nonumber\\
 = & \int_{0}^{h^\ast} x^r\,\biggl\{\left(1+x\right)|\alpha|^2
 +\left(1-x\right)|\beta|^2
 -\frac{2\rho_0\Re(\alpha\overline{\beta}e^{i\nu})}{\rho}\,x\biggr\}\,\frac{1}{2\pi}\cdot\frac{dk_{-}(x)}{dx}\,dx\nonumber\\
 & +\int_{h^\ast}^{0} x^r\,\biggl\{\left(1+x\right)|\alpha|^2
 +\left(1-x\right)|\beta|^2
 -\frac{2\rho_0\Re(\alpha\overline{\beta}e^{i\nu})}{\rho}\,x\biggr\}\,\frac{1}{2\pi}\cdot\frac{dk_{+}(x)}{dx}\,dx\nonumber\\
 & + \int_{0}^{h^\ast} (-x)^r\,\biggl\{\left(1-x\right)|\alpha|^2
 +\left(1+x\right)|\beta|^2
 +\frac{2\rho_0\Re(\alpha\overline{\beta}e^{i\nu})}{\rho}\,x\biggr\}\,\frac{1}{2\pi}\cdot\frac{dk_{-}(x)}{dx}\,dx\nonumber\\
 & + \int_{h^\ast}^{0} (-x)^r\,\biggl\{\left(1-x\right)|\alpha|^2
 +\left(1+x\right)|\beta|^2
 +\frac{2\rho_0\Re(\alpha\overline{\beta}e^{i\nu})}{\rho}\,x\biggr\}\,\frac{1}{2\pi}\cdot\frac{dk_{+}(x)}{dx}\,dx\nonumber\\
 = & \int_{0}^{h^\ast} x^r\,\biggl\{\left(1+x\right)|\alpha|^2
 +\left(1-x\right)|\beta|^2
 -\frac{2\rho_0\Re(\alpha\overline{\beta}e^{i\nu})}{\rho}\,x\biggr\}\,\frac{1}{2\pi}\left(\frac{dk_{-}(x)}{dx}-\frac{dk_{+}(x)}{dx}\right)\,dx\nonumber\\
 & + \int_{0}^{h^\ast} (-x)^r\,\biggl\{\left(1-x\right)|\alpha|^2
 +\left(1+x\right)|\beta|^2
 +\frac{2\rho_0\Re(\alpha\overline{\beta}e^{i\nu})}{\rho}\,x\biggr\}\,\frac{1}{2\pi}\left(\frac{dk_{-}(x)}{dx}-\frac{dk_{+}(x)}{dx}\right)\,dx\nonumber\\
 = & \int_{0}^{h^\ast} x^r\,\biggl\{\left(1+x\right)|\alpha|^2
 +\left(1-x\right)|\beta|^2
 -\frac{2\rho_0\Re(\alpha\overline{\beta}e^{i\nu})}{\rho}\,x\biggr\}\,\frac{1}{2\pi}\left(\frac{dk_{-}(x)}{dx}-\frac{dk_{+}(x)}{dx}\right)\,dx\nonumber\\
 & + \int_{-h^\ast}^{0} x^r\,\biggl\{\left(1+x\right)|\alpha|^2
 +\left(1-x\right)|\beta|^2
 -\frac{2\rho_0\Re(\alpha\overline{\beta}e^{i\nu})}{\rho}\,x\biggr\}\,\frac{1}{2\pi}\left(\frac{dk_{-}(x)}{dx}-\frac{dk_{+}(x)}{dx}\right)\,dx\nonumber\\
 = & \int_{-h^\ast}^{h^\ast} x^r\,\biggl\{\left(1+x\right)|\alpha|^2
 +\left(1-x\right)|\beta|^2
 -\frac{2\rho_0\Re(\alpha\overline{\beta}e^{i\nu})}{\rho}\,x\biggr\}\,\frac{1}{2\pi}\left(\frac{dk_{-}(x)}{dx}-\frac{dk_{+}(x)}{dx}\right)\,dx\nonumber\\
 = & \int_{-h^\ast}^{h^\ast} x^r\,\biggl\{1+\left(|\alpha|^2-|\beta|^2-\frac{2\rho_0\Re(\alpha\overline{\beta}e^{i\nu})}{\rho}\right)\,x\biggr\}\,\frac{1}{2\pi}\left(\frac{dk_{-}(x)}{dx}-\frac{dk_{+}(x)}{dx}\right)\,dx,\label{eq:transform_moment}
\end{align}
where
\begin{align}
 k_{\pm}(x)=&\arcsin\left(\frac{-\rho\rho_0(\sin\nu)x^2\pm\sqrt{\xi(x)}}{\rho\rho_0(1-x^2)}\right),\\
 \frac{dk_{\pm}(x)}{dx}=&\mp\frac{\sqrt{\eta_{\pm}(x)}}{(1-x^2)\sqrt{\xi(x)}},\\
 \xi(x)=&(\rho^2-x^2)(\rho_0^2-x^2)-\rho^2\rho_0^2(\cos^2\nu) x^2,\\
 \eta_{\pm}(x)=& 1-\rho^2\rho_0^2(1+\sin^2\nu)-\left(1-\rho^2\rho_0^2\cos^2\nu\right)\,x^2\pm 2\rho\rho_0\,(\sin\nu)\sqrt{\xi(x)}.
\end{align}
We have remembered $|\alpha|^2+|\beta|^2=1$ at the final line in Eq.~\eqref{eq:transform_moment}.
With another form of $\Re(\alpha\overline{\beta}e^{i\nu})=\Re(\alpha\overline{\beta})\cos\nu-\Im(\alpha\overline{\beta})\sin\nu$, from which
\begin{align}
 &1+\left(|\alpha|^2-|\beta|^2-\frac{2\rho_0\Re(\alpha\overline{\beta}e^{i\nu})}{\rho}\right)\,x\nonumber\\
 =& 1+\Biggl\{\left|\alpha\right|^2-\left|\beta\right|^2-\frac{2\rho_0}{\rho}\Bigl(\Re(\alpha\overline{\beta})\cos\nu-\Im(\alpha\overline{\beta})\sin\nu\Bigr)\Biggr\}\,x\nonumber\\
 =& \gamma(x)
\end{align}
follows, we find
\begin{equation}
  \sum_{j=1}^2 \int_{-\pi}^{\pi} \left(\frac{i\,\lambda'_j(k)}{\lambda_j(k)}\right)^r \Bigl|\braket{v_j(k)|\tilde\phi}\Bigr|^2 \frac{dk}{2\pi}
 = \int_{-h^\ast}^{h^\ast} x^r \cdot \frac{\sqrt{\eta_{+}(x)} + \sqrt{\eta_{-}(x)}}{2\pi(1-x^2)\sqrt{\xi(x)}}\,\gamma(x)\,dx.\label{eq:transform_moment_2}
\end{equation}
Equation~\eqref{eq:transform_moment_2} is rewritten with the indicator function $I_{(-h^\ast, h^\ast)}(x)$, which was defined in Eq.~\eqref{eq:indicator}, as the following form,
\begin{equation}
  \sum_{j=1}^2 \int_{-\pi}^{\pi} \left(\frac{i\,\lambda'_j(k)}{\lambda_j(k)}\right)^r \Bigl|\braket{v_j(k)|\tilde\phi}\Bigr|^2 \frac{dk}{2\pi}
 = \int_{-\infty}^{\infty} x^r \cdot \frac{\sqrt{\eta_{+}(x)} + \sqrt{\eta_{-}(x)}}{2\pi(1-x^2)\sqrt{\xi(x)}}\,\gamma(x)\, I_{(-h^\ast, h^\ast)}(x)\,dx.
\end{equation}
Therefore, returning from the Fourier space, we reach another form of the limit
\begin{equation}
 \lim_{t\to\infty}\mathbb{E}\left[\left(\frac{X_t}{t}\right)^r\right]
  =\int_{-\infty}^{\infty} x^r \cdot \frac{\sqrt{\eta_{+}(x)} + \sqrt{\eta_{-}(x)}}{2\pi(1-x^2)\sqrt{\xi(x)}}\,\gamma(x)\, I_{(-h^\ast, h^\ast)}(x)\,dx,\label{eq:lim_E}.
\end{equation}
This convergence of the $r$-th moments shown in Eq.~\eqref{eq:lim_E} guarantees the statement of Theorem~\ref{th:limit_CMV}.

\clearpage

\appendix{~$\ket{\Psi_{t+1}}=V\ket{\Psi_t}$}

If we do not use the operation $U_f$, the operation onto the Fourier transform $\ket{\hat\psi_t(k)}$ contains a 2-step evolution of a standard coined quantum walk,
\begin{equation}
 \hat{U}(k)
  =R\left(\frac{\nu}{2}\right)
  \left\{R\left(-\frac{k}{2}\right)
  \begin{bmatrix}
   \rho_0 & -\rho\\
   \rho & \rho_0
  \end{bmatrix}\right\}^2
  R\left(-\frac{\nu}{2}\right),
\end{equation}
from which
\begin{align}
 \ket{\hat\psi_t(k)}
 =&\hat{U}(k)^t\ket{\hat\psi_0(k)}\nonumber\\
 =&R\left(\frac{\nu}{2}\right)
  \left\{R\left(-\frac{k}{2}\right)
  \begin{bmatrix}
   \rho_0 & -\rho\\
   \rho & \rho_0
  \end{bmatrix}\right\}^{2t}
  R\left(-\frac{\nu}{2}\right)\ket{\phi}.
\end{align}
Note that
\begin{align}
 \left\{R\left(-\frac{k}{2}\right)
  \begin{bmatrix}
   \rho_0 & -\rho\\
   \rho & \rho_0
  \end{bmatrix}\right\}^2
 =&\begin{bmatrix}
    \rho_0^2 e^{-ik}-\rho^2 & -\rho\rho_0 e^{-ik}-\rho\rho_0\\
    \rho\rho_0 e^{ik}+\rho\rho_0 & \rho_0^2 e^{ik}-\rho^2
   \end{bmatrix},\\
 R\left(-\frac{\nu}{2}\right)\ket{\phi}
  =&\begin{bmatrix}
    e^{-i\nu/2}\alpha\\[1mm]
    e^{i\nu/2}\beta
   \end{bmatrix},
\end{align}
and $|e^{-i\nu/2}\alpha|^2+|e^{i\nu/2}\beta|^2=|\alpha|^2+|\beta|^2=1$.
The walker is, hence, observed at position $x$ at time $t$ with probability
\begin{align}
 \mathbb{P}(X_t=x)
 =&\left|\left|\,  R\left(\frac{\nu}{2}\right)\int_{-\pi}^\pi e^{ikx}
  \left\{R\left(-\frac{k}{2}\right)
  \begin{bmatrix}
   \rho_0 & -\rho\\
   \rho & \rho_0
  \end{bmatrix}\right\}^{2t}
  R\left(-\frac{\nu}{2}\right)\ket{\phi}\,\frac{dk}{2\pi}\,\right|\right|^2\nonumber\\
 =&\left|\left|\, \int_{-\pi}^\pi e^{ikx}
  \left\{R\left(-\frac{k}{2}\right)
  \begin{bmatrix}
   \rho_0 & -\rho\\
   \rho & \rho_0
  \end{bmatrix}\right\}^{2t}
  R\left(-\frac{\nu}{2}\right)\ket{\phi}\,\frac{dk}{2\pi}\,\right|\right|^2.\label{eq:app_prob}
\end{align}
The initial state of the Fourier transform should be reconsidered as $R\left(-\nu/2\right)\ket{\phi}$ in Eq.~\eqref{eq:app_prob}.
By a similar Fourier analysis used for quantum walks in~\cite{Machida2016a}, the walker converges in distribution as $t\to\infty$.  

\bigskip

For any real number $x$, we have
\begin{align}
 \lim_{t\to\infty}\mathbb{P}\left(\frac{X_t}{t}\leq x\right)
 =&\int_{-\infty}^x \frac{\rho}{\pi(1-y^2)\sqrt{\rho_0^2-y^2}}\,\Delta(y)\,I_{(-\rho_0,\, \rho_0\,)}(y)\,dy,
\end{align}
where
\begin{align}
 \Delta(x)=& 1+\biggl\{|\alpha|^2+|\beta|^2-\frac{2\rho}{\rho_0}\left(\Re(\alpha\overline{\beta})\cos\nu+\Im(\alpha\overline{\beta})\sin\nu\right)\biggr\}x,\\
 I_{(-\rho_0,\, \rho_0)}(x)=&\left\{\begin{array}{cl}
			      1&(x\in (-\rho_0,\, \rho_0\,))\\
				     0&(x\notin (-\rho_0,\, \rho_0\,))
				    \end{array}\right..
\end{align}

\bigskip

\begin{figure}[h]
\begin{center}
\begin{minipage}{50mm}
  \begin{center}
  (a) $(\alpha, \beta) = (1/\sqrt{2}\,,i/\sqrt{2}\,)$
  \end{center}
 \end{minipage}
 \begin{minipage}{50mm}
  \begin{center}
  (b) $(\alpha, \beta) = (1,0)$
  \end{center}
 \end{minipage}\\
 \bigskip
 \begin{minipage}{50mm}
  \begin{center}
   \includegraphics[scale=0.4]{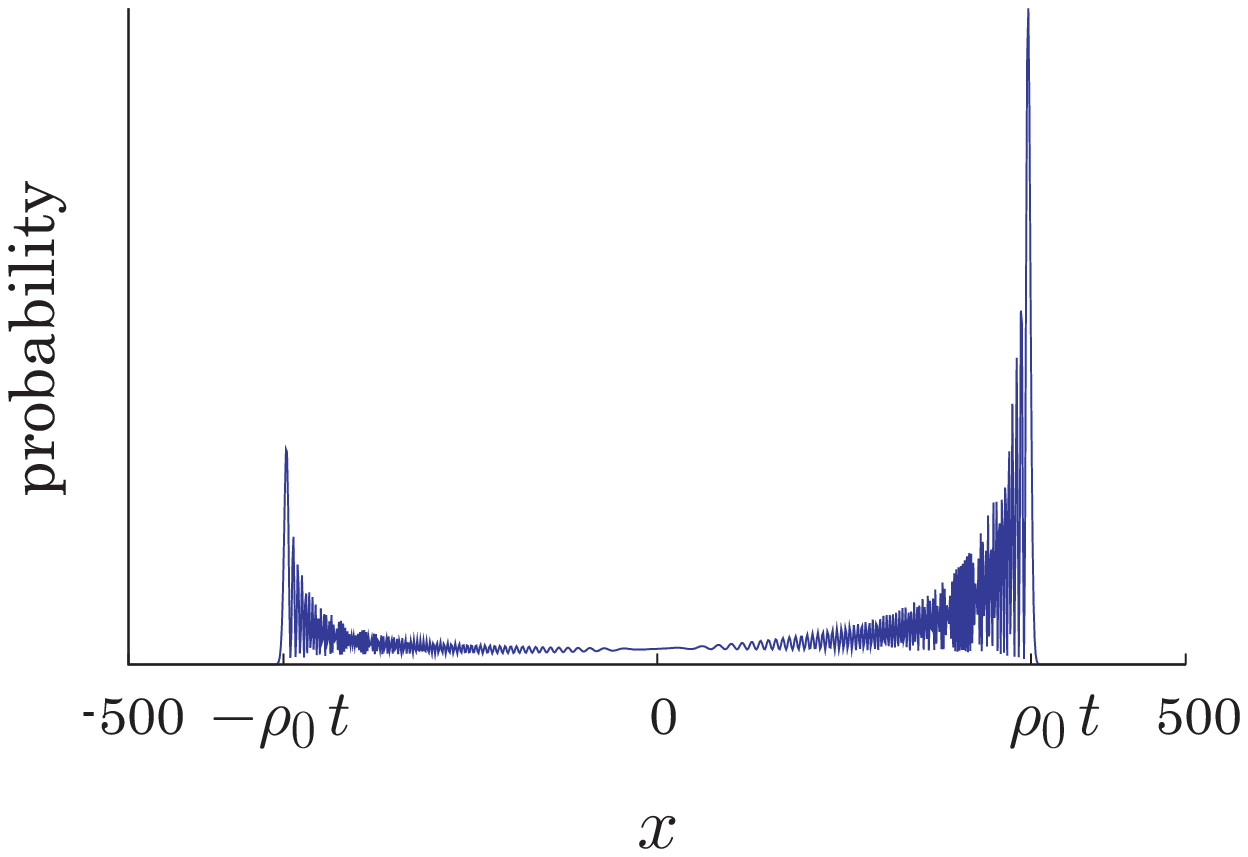}\\[2mm]
  (a)--1
  \end{center}
 \end{minipage}
 \begin{minipage}{50mm}
  \begin{center}
   \includegraphics[scale=0.4]{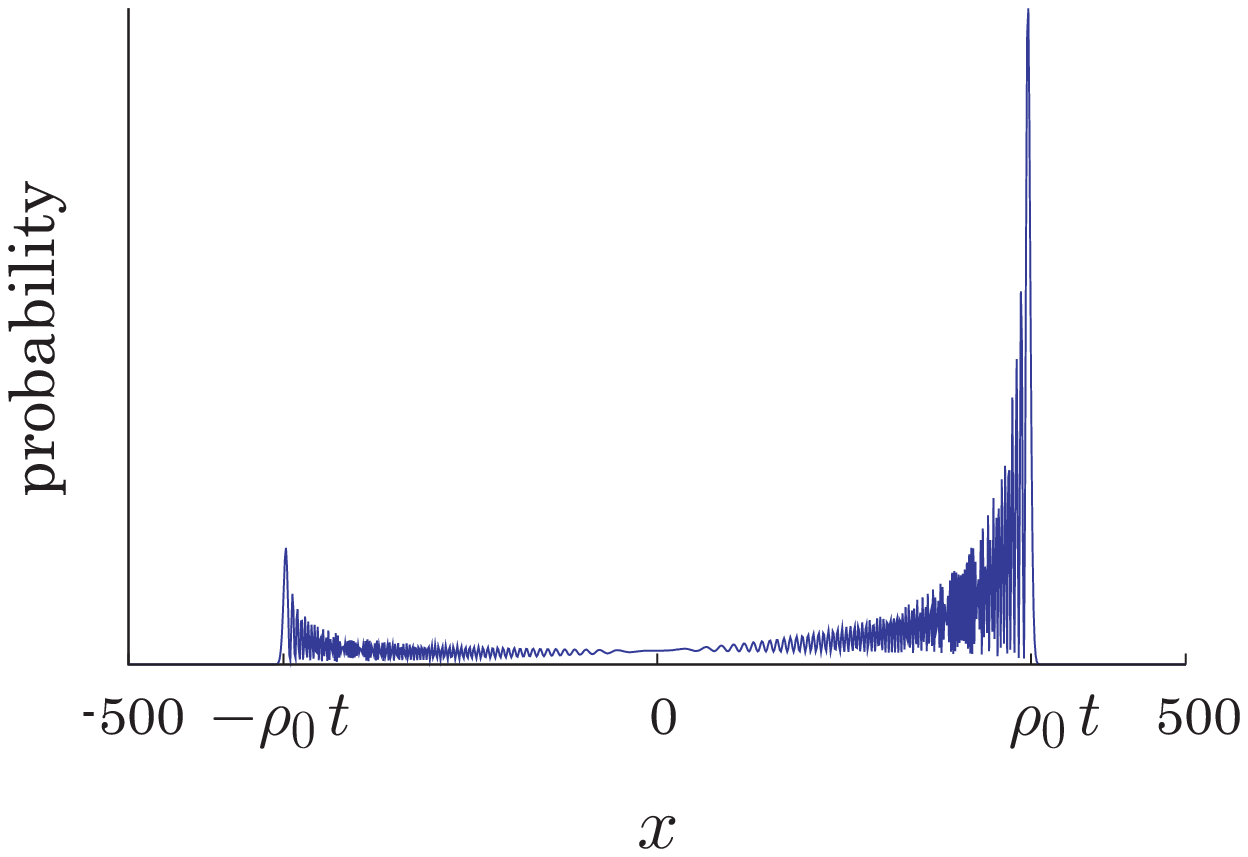}\\[2mm]
  (b)--1
  \end{center}
 \end{minipage}
 \bigskip
 \begin{minipage}{50mm}
  \begin{center}
   \includegraphics[scale=0.4]{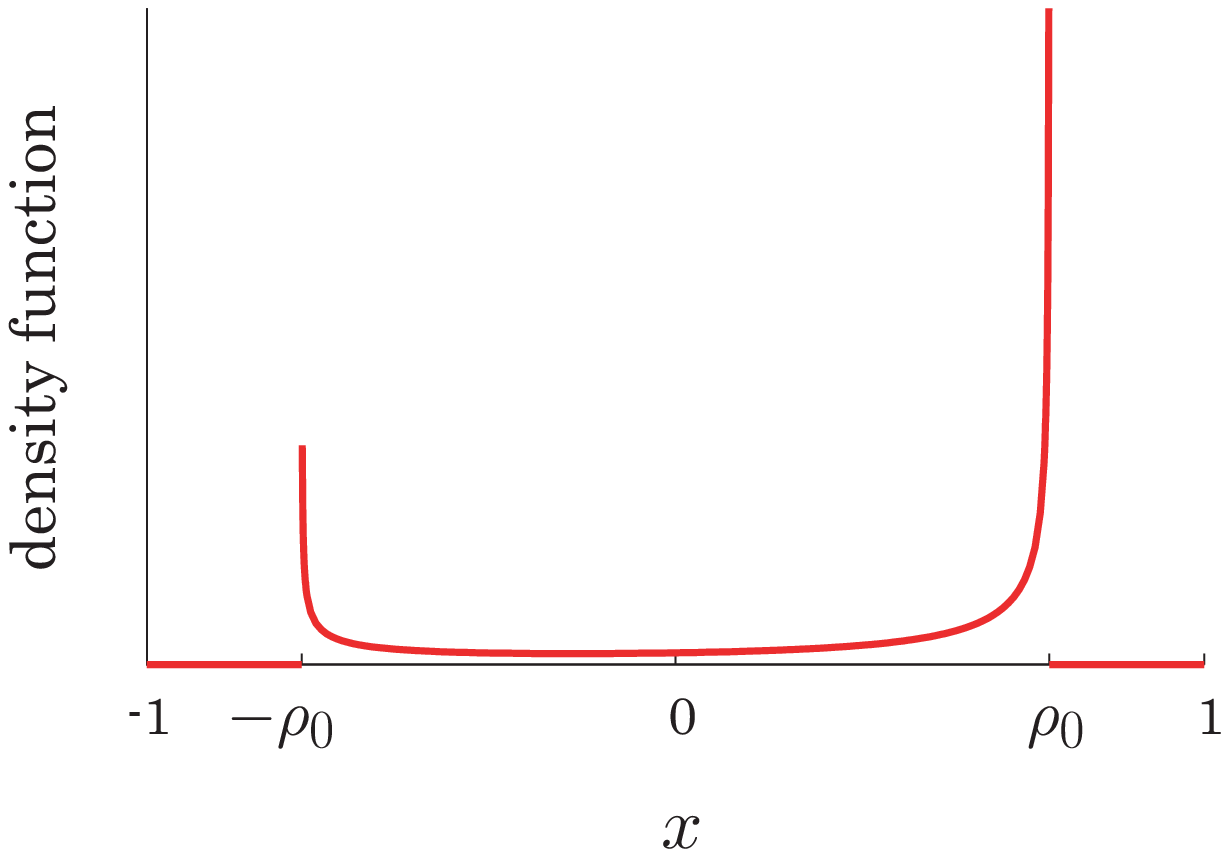}\\[2mm]
  (a)--2
  \end{center}
 \end{minipage}
 \begin{minipage}{50mm}
  \begin{center}
   \includegraphics[scale=0.4]{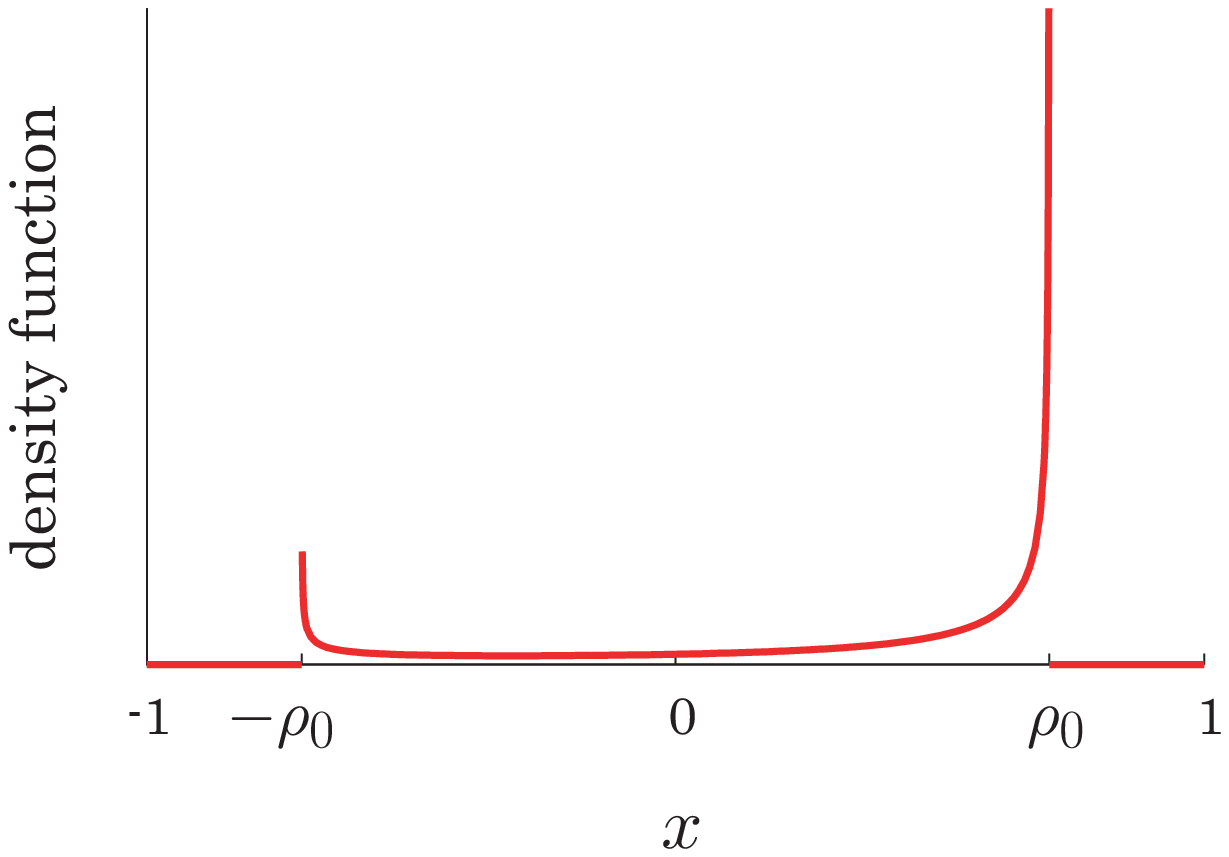}\\[2mm]
  (b)--2
  \end{center}
 \end{minipage}
 \bigskip
\begin{minipage}{50mm}
  \begin{center}
   \includegraphics[scale=0.4]{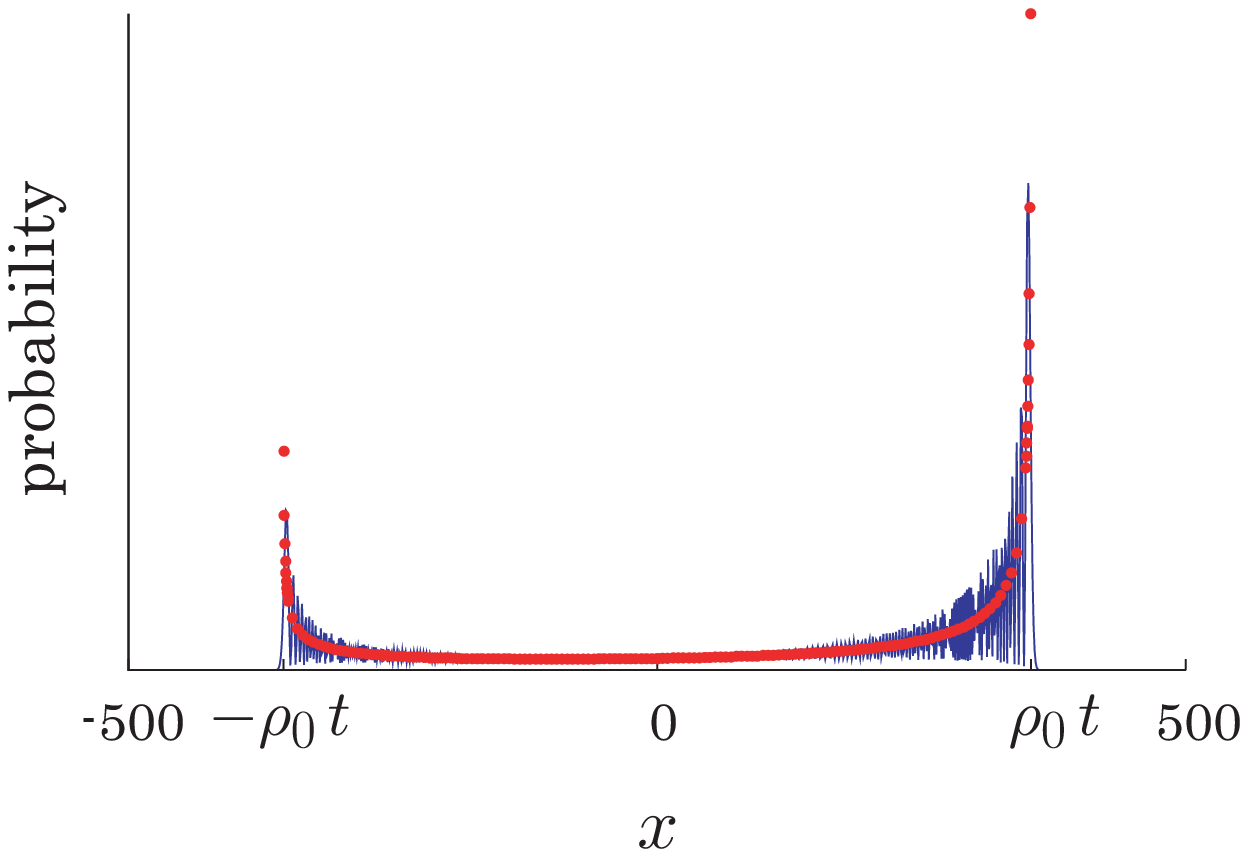}\\[2mm]
  (a)--3
  \end{center}
 \end{minipage}
 \begin{minipage}{50mm}
  \begin{center}
   \includegraphics[scale=0.4]{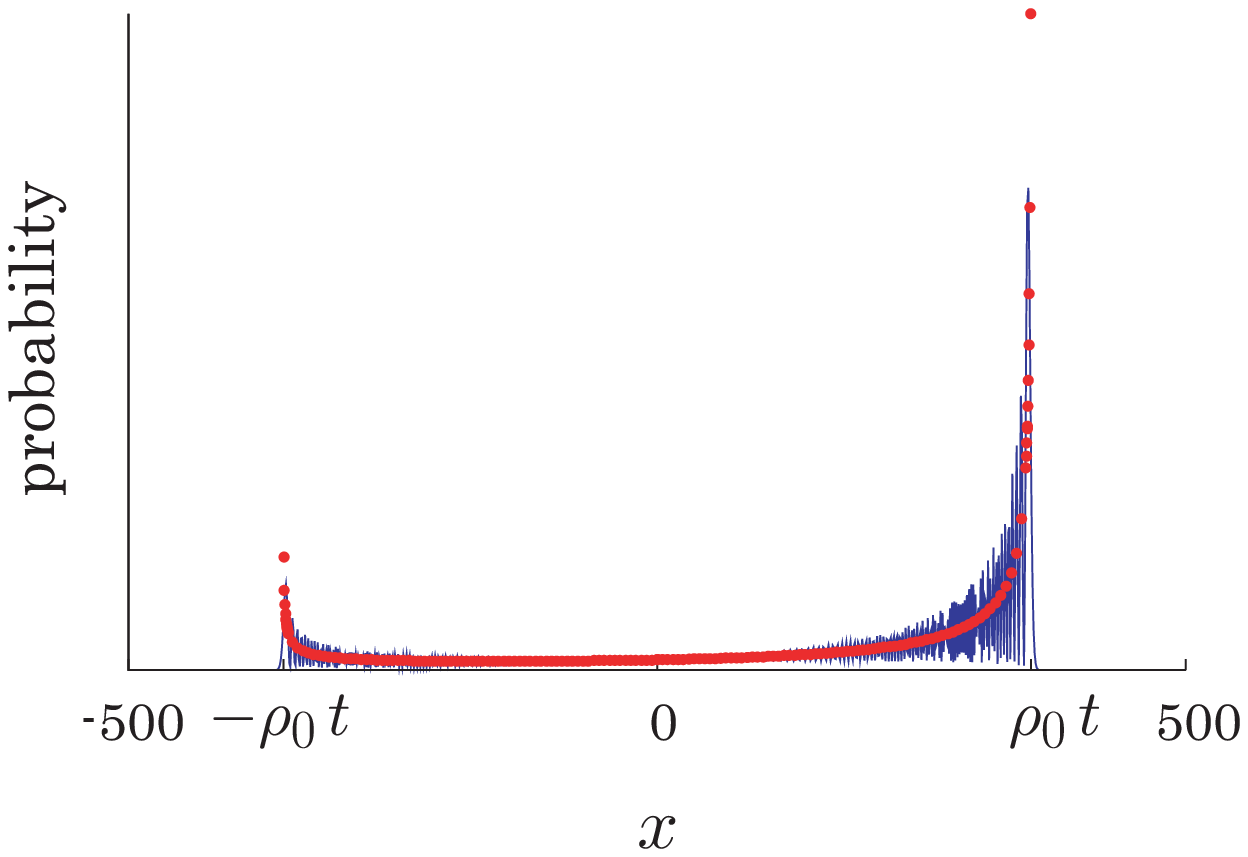}\\[2mm]
  (b)--3
  \end{center}
 \end{minipage}
\fcaption{$\rho=1/\sqrt{2}, \nu=\pi/4$ :
The blue lines represent the probability distribution $\mathbb{P}(X_t=x)$ at time $t=500$ ((a)--1, (b)--1) and the red lines represent the limit density function ((a)--2, (b)--2). 
In (a)--3 and (b)--3, we confirm that the approximation (red points) obtained from the limit density function reproduces the features of the probability distribution as time $t$ becomes large enough.
The walker launches with the localized initial state at the origin, $\ket{\Psi_0}=\ket{0}\otimes (\alpha\ket{0}+\beta\ket{1})$.}
\label{fig:3}
\end{center}
\end{figure}

This result for the coined quantum walk is already contained in~\cite{Konno2008}.
The proof for this limit theorem is omitted in this paper because of the similarity to the computation written in~\cite{Machida2016a}.
Figure~\ref{fig:3} shows the behavior of the quantum walk when the operation lacks the unitary matrix $U_f$.

\clearpage



\begin{thebibliography}{1}

\bibitem{AharonovDavidovichZagury1993}
Y. Aharonov, L. Davidovich and N. Zagury
\newblock (1993), \textit{Quantum random walks}, Phys. Rev. A, 48(2), pp.
  1687--1690.

\bibitem{Konno2008}
N. Konno
\newblock (2008), In:
\newblock \textit{Quantum Walks}. Vol. 1954 of Lecture Notes in Mathematics,
  Springer-Verlag
\newblock Heidelberg, pp.  309--452.

\bibitem{GrimmettJansonScudo2004}
G. Grimmett, S. Janson and P.F. Scudo
\newblock (2004), \textit{Weak limits for quantum random walks}, Phys. Rev. E,
  69(2),  026119.

\bibitem{MachidaKonno2010}
T. Machida and N. Konno
\newblock (2010), \textit{Limit theorem for a time-dependent coined quantum
  walk on the line}, F. Peper et al. (Eds.): IWNC 2009, Proceedings in
  Information and Communications Technology, 2, pp.  226--235.

\bibitem{CanteroMoralGrunbaumVelazquez2010}
M.J. Cantero, L. Moral, F.A. Gr{\"u}nbaum and L. Vel{\'a}zquez
\newblock (2010), \textit{Matrix-valued szeg{\H{o}} polynomials and quantum
  random walks}, Communications on Pure and Applied Mathematics, 63(4), pp.
  464--507.

\bibitem{Machida2016a}
T. Machida
\newblock (2016), In:
\newblock \textit{Research Advances in Quantum Dynamics}. InTech
\newblock pp.  27--51.

\end{thebibliography}
\end{document}